\documentclass[11pt]{cernrep}
\usepackage{epsfig}
\usepackage{graphicx}
\usepackage{cite,./mcite}

\newcommand{\leqsim}{\,\raisebox{-0.6ex}{$\buildrel < \over \sim$}\,}
\newcommand{\geqsim}{\,\raisebox{-0.6ex}{$\buildrel > \over \sim$}\,}

\newcommand{\asmz}{\alpha_s(M_Z^2)}
\newcommand{\msbar}{\mbox{$\overline{\rm{MS}}$}\ }

\begin{document}
\title{Hadron Structure from Inclusive and Exclusive Cross-Sections in ep scattering \\ }
\author{Amanda Cooper-Sarkar}
\institute{Oxford University}
\maketitle
\begin{abstract}
The contribution of both inclusive and exclusive cross-section data from HERA to our 
knowledge of parton distribution functions is reviewed and future prospects are outlined. 
\end{abstract}

\section{Introduction}
\label{sec:intro}
The kinematics
of lepton hadron scattering is described in terms of the variables $Q^2$, the
invariant mass of the exchanged vector boson, Bjorken $x$, the fraction
of the momentum of the incoming nucleon taken by the struck quark (in the 
quark-parton model), and $y$ which measures the energy transfer between the
lepton and hadron systems.
The differential cross-section for the neutral current (NC) process is given in 
terms of the structure functions by
\[
\frac {d^2\sigma(e^{\pm}p) } {dxdQ^2} =  \frac {2\pi\alpha^2} {Q^4 x}
\left[Y_+\,F_2(x,Q^2) - y^2 \,F_L(x,Q^2)
\mp Y_-\, xF_3(x,Q^2) \right],
\]
where $\displaystyle Y_\pm=1\pm(1-y)^2$. 
The structure functions $F_2$ and $xF_3$ are 
directly related to quark distributions, and their
$Q^2$ dependence, or scaling violation, 
is predicted by pQCD. For low $x$, $x \leq 10^{-2}$, $F_2$ 
is sea quark dominated, but its $Q^2$ evolution is controlled by
the gluon contribution, such that HERA data provide 
crucial information on low-$x$ sea-quark and gluon distributions.
At high $Q^2$, the structure function $xF_3$ becomes increasingly 
important, and gives information on valence quark distributions. 
The charged current (CC) interactions also
enable us to separate the flavour of the valence distributions 
at high-$x$, since their (LO) cross-sections are given by, 
\[
\frac {d^2\sigma(e^+ p) } {dxdQ^2} = \frac {G_F^2 M_W^4} {(Q^2 +M_W^2)^2 2\pi x}
x\left[(\bar{u}+\bar{c}) + (1 - y)^2 (d + s) \right],
\]
\[
\frac {d^2\sigma(e^- p) } {dxdQ^2} = \frac {G_F^2 M_W^4} {(Q^2 +M_W^2)^2 2\pi x}
x\left[(u + c) + (1 - y)^2 (\bar{d} + \bar{s}) \right].
\]

Parton Density Function (PDF) determinations are usually global 
fits~\cite{mrst,cteq,zeus-s}, which use fixed target 
DIS data as well as HERA data. In such analyses, the high statistics HERA NC 
$e^+p$ data have determined the low-$x$ sea and 
gluon distributions, whereas the fixed target data have determined 
the valence distributions. Now that high-$Q^2$ HERA data on NC and CC
 $e^+p$ and $e^-p$ inclusive double 
differential cross-sections are available, PDF fits can be made to HERA 
data alone, since the HERA high $Q^2$ cross-section 
data can be used to determine the valence distributions. This has the 
advantage that it eliminates the need for heavy target corrections, which 
must be applied to the $\nu$-Fe and $\mu D$ fixed target data. Furthermore
there is no need to assume isospin symmetry, i.e. that $d$ in the 
proton is the same as $u$ in the neutron, 
since the $d$ distribution can be obtained directly from CC $e^+p$ data. 

The H1 and ZEUS collaborations have both used their data to make PDF 
fits~\cite{zeusj,*h1}. In Section~\ref{sec:compare} we review the published PDF analyses 
paying particular attention to the treatment of correlated systematic errors. In
Section~\ref{sec:combine} we present the preliminary results of a combination of 
ZEUS and H1 data. In Section~\ref{sec:jets} we discuss the improvement in our knowledge 
of the gluon PDF, which comes from the addition of jet data to the PDF fits, and we 
present the measurements of $\alpha_s$ which have been made using HERA jet data. 
In Section~\ref{sec:heraii} we present preliminary fits using HERA-II data and in 
Section~\ref{sec:future} we conclude by looking at the propsects for the future. 

\section{Comparing ZEUS and H1 published PDF analyses}
\label{sec:compare}
Full details of the analyses are given in the 
relevant publications, in this contribution we examine the differences in 
the two analyses, recapping only salient details. 
For both HERA analyses the QCD predictions for the structure functions 
are obtained by solving the DGLAP evolution equations~\cite{ap,*gl,*l,*d} 
at NLO in the \msbar scheme with the
renormalisation and factorization scales chosen to be $Q^2$. 
These equations yield the PDFs
 at all values of $Q^2$ provided they
are input as functions of $x$ at some input scale $Q^2_0$. 
The resulting PDFs are then convoluted with coefficient functions, to 
give the structure functions which enter into the expressions for the 
cross-sections. For the ZEUS analysis, the coefficient functions are 
calculated using the 
general-mass variable flavour number scheme of Roberts and Thorne~\cite{hq}.
For the H1 analysis, the zero-mass variable flavour number scheme is used. 

The HERA data are all in a kinematic region where there is no
sensitivity to target mass and higher 
twist contributions, but a minimum $Q^2$ cut must be imposed 
to remain in the kinematic region where
perturbative QCD should be applicable. For ZEUS this is $Q^2 > 2.5$~GeV$^2$, 
and for H1 it is $Q^2 > 3.5$~GeV$^2$. Both collaborations have included the 
sensitivity to this cut as part of their model uncertainties.

In the ZEUS analysis (called the ZEUS-JETS fit), the PDFs for $u$ valence, $xu_v(x)$,  $d$ valence, $xd_v(x)$, 
total sea, $xS(x)$, the 
gluon, $xg(x)$, and the difference between the $d$ and $u$
contributions to the sea, $x(\bar{d}-\bar{u})$, are each parametrized  
by the form 
\begin{equation}
  p_1 x^{p_2} (1-x)^{p_3} P(x),
\label{eqn:pdf}
\end{equation}
where $P(x) = 1 +p_4 x$, at $Q^2_0 = 7$GeV$^2$. The total sea 
$xS=2x(\bar{u} +\bar{d} +\bar{s}+ \bar{c} +\bar{b})$, where 
$\bar{q}=q_{sea}$ for each flavour, $u=u_v+u_{sea}, d=d_v+d_{sea}$ and 
$q=q_{sea}$ for all other flavours. 
The flavour structure of the light quark sea 
allows for the violation of the Gottfried sum rule. However, there is no 
information on the shape of the $\bar{d}-\bar{u}$ distribution in a fit 
to HERA data alone and so this distribution has its shape fixed consistent 
with the Drell-Yan data and its normalisation consistent 
with the size of the Gottfried sum-rule violation. 
A suppression of the strange sea with respect to the non-strange sea 
of a factor of 2 at $Q^2_0$, is also imposed
consistent with neutrino induced dimuon data from CCFR. 
Parameters are further restricted as follows.
The normalisation parameters, $p_1$, for the $d$ and $u$ valence and for the 
gluon are constrained to impose the number sum-rules and momentum sum-rule. 
The $p_2$ parameter which constrains the low-$x$ behaviour of the $u$ and $d$ 
valence distributions is set equal, 
since there is no information to constrain any difference. 
In the present fits to HERA-I data it is also necessary to constrain 
the high-$x$ sea and gluon shapes, because HERA-I data do not have high 
statistics at large-$x$, in the region where these distributions are small.
The sea shape has been restricted by setting $p_4=0$ for the sea, 
but the gluon shape is constrained by including data on jet production
 in the PDF fit, as discussed in Sec.~\ref{sec:jets}. 
Finally the ZEUS analysis has 11 free PDF parameters. 
ZEUS have included reasonable variations of 
these assumptions about the input parametrization 
in their analysis of model uncertainties. 
The strong coupling constant was fixed to $\asmz =  0.118$~\cite{lepalf}.
Full account has been taken of correlated experimental 
systematic errors by the Offset Method, 
as described in ref~\cite{zeus-s,durham}.

For the H1 analysis (called the H1 2000 PDF fit), the value of $Q^2_0 = 4$GeV$^2$, and 
the choice of quark distributions which are 
parametrized is different. The quarks are considered as $u$-type and $d$-type
with different parametrizations for, $xU= x(u_v+u_{sea} + c)$, 
$xD= x(d_v +d_{sea} + s)$, $x\bar{U}=x(\bar{u}+\bar{c})$ and 
$x\bar{D}=x(\bar{d}+\bar{s})$, with $q_{sea}=\bar{q}$, as 
usual, and the the form of the quark and gluon parametrizations
given by Eq.~\ref{eqn:pdf}. For $x\bar{D}$ and $x\bar{U}$ the polynomial, 
$P(x)=1.0$,
for the gluon and $xD$, $P(x)= (1+p_4 x)$, and for $xU$, 
$P(x)= (1 +p_4 x +p_5 x^3)$. The parametrization is then further restricted 
as follows.
Since the valence distributions must vanish as $x \to 0$, 
the low-$x$ parameters, $p_1$
 and $p_2$ are set equal for $xU$ and $x\bar{U}$, and for $xD$ and 
$x\bar{D}$. Since there is no information on the flavour structure of the sea 
it is 
also necessary to set $p_2$ equal for $x\bar{U}$ and $x\bar{D}$. 
The normalisation, $p_1$, of the gluon is determined from the momentum 
sum-rule and the $p_4$ parameters for $xU$ and $xD$ are determined from the 
valence number sum-rules.
Assuming that the strange and charm quark distributions can be expressed as 
$x$ independent fractions, $f_s$ and $f_c$, of the $d$ and $u$ type sea, 
gives the further constraint $p_1(\bar{U})=p_1(\bar{D}) (1-f_s)/(1-f_c)$. 
Finally there are 10 free parameters. H1 have also included reasonable 
variations of 
these assumptions in their analysis of model uncertainties. 
The strong coupling constant was fixed to $\asmz =  0.1185$ and this is 
sufficiently similar to the ZEUS choice that we can rule it out as a cause of
any significant difference. 
Full account has been taken of correlated experimental 
systematic errors by the Hessian Method, see ref.~\cite{durham}. 
 
The different treatments of correlated experimental systematic errors deserves a little 
more dicussion since modern deep inelastic scattering experiments
have very small statistical uncertainties, so that the contribution of 
systematic uncertainties becomes dominant and consideration of 
point to point correlations between systematic uncertainties is essential.

For both ZEUS and H1 analyses
the formulation of the $\chi^2$ including correlated systematic uncertainties
 is constructed as follows. The correlated uncertainties
are included in the theoretical prediction, $F_i(p,s)$, such that
\[ 
F_i(p,s) = F_i^{\rm NLOQCD}(p) + 
\sum_{\lambda} s_{\lambda} \Delta^{\rm sys}_{i\lambda}
\]
where, $F_i^{\rm NLOQCD}(p)$, represents the prediction 
from NLO QCD in terms of the theoretical parameters $p$,
and the parameters $s_\lambda$ represent independent variables 
for each source of
 systematic uncertainty. They have zero mean and unit variance by construction.
The symbol 
$\Delta^{\rm sys}_{i\lambda}$ represents the one standard deviation correlated 
systematic error on data point $i$ due to correlated error 
source $\lambda$.
The $\chi^2$ is then formulated as 
\begin{equation}
\chi^2 = \sum_i \frac{\left[ F_i(p,s)-F_i(\rm meas) \right]^2}{\sigma_i^2} + \sum_\lambda s^2_\lambda 
\label{eq:chi2}
\end{equation}
where, $F_i(\rm meas)$, represents a measured data point and the symbol 
$\sigma_i$ represents the one standard deviation uncorrelated 
error on data point $i$, from both statistical and systematic sources. 
The experiments use this $\chi^2$ in different ways. ZEUS uses the Offset 
method and H1 uses the Hessian method.
 
Traditionally, experimentalists have used `Offset' methods to account for
correlated systematic errors. The $\chi^2$ is formulated without any terms
due to correlated systematic errors ($s_\lambda=0$ in Eq.~\ref{eq:chi2}) for
evaluation of the central values of the fit parameters. 
However, the data points are then offset to account for each 
source of systematic error in turn 
(i.e. set $s_\lambda = + 1$ and then $s_\lambda = -1$ for each source 
$\lambda$) 
and a new fit is performed for each of these
variations. The resulting deviations of the theoretical parameters 
from their  central  values are added in 
quadrature. (Positive and  negative deviations are added 
in quadrature separately.)
This procedure gives fitted theoretical predictions which are as close as 
possible to the central values of the published data. It does not use 
the full statistical power of the fit to improve the estimates of $s_\lambda$, 
and thus it is a more conservative method of error estimation  
than the Hessian method.

The Hessian method is an alternative procedure in which the systematic
uncertainty parameters $s_\lambda$ are allowed to vary in the main fit 
when determining the values of the theoretical parameters. 
Effectively, the theoretical prediction is not fitted
to the central values of the published experimental data, but  
these data points are allowed to move
collectively, according to their correlated systematic uncertainties.
 The theoretical prediction determines the 
optimal settings for correlated systematic shifts of experimental data 
points 
such that the most consistent fit to all data sets is obtained. Thus, 
in a global fit, systematic shifts in 
one experiment are correlated to those in another experiment by the fit.
In essence one is allowing the theory to calibrate the detectors. This requires
confidence in the theory, but more significantly, it requires confidence
in the many model choices (such as the parametrization at $Q^2_0$) 
which go into setting the boundary conditions for the theory . 

\begin{figure}[tbp]
\vspace{-2.0cm} 
\centerline{
\epsfig{figure=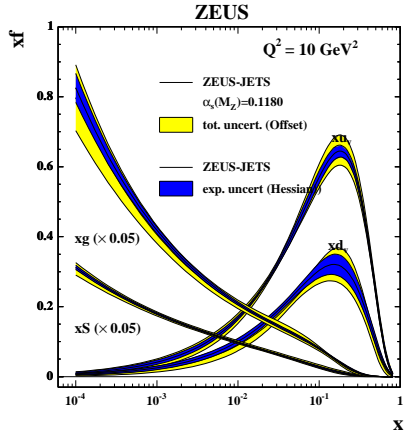,height=6cm}
}
\caption {PDFs at $Q^2=10$GeV$^2$, for the ZEUS analysis  
comparing the Offset and the Hessian methods.
}
\label{fig:offhess}
\end{figure}

To compare these two methods the ZEUS analysis has been 
performed using the Hessian method as well as the 
Offset method and Fig.~\ref{fig:offhess} compares the PDFs, and their uncertainties, 
using these two methods. 
The central values of the different methods are in good agreement but 
the use of the Hessian method results in smaller uncertainties, for a 
the standard set of model assumptions.  
However, model uncertainties are more significant for the Hessian method than 
for the Offset method. The PDF
parameters obtained for different model choices can differ 
by much more than their experimental uncertainties, 
because each model choice can result in somewhat 
different values of the systematic uncertainty parameters, 
$s_\lambda$, and 
thus a different estimate of the shifted positions of the data points. 
This 
results in a larger spread of model uncertainty than is found in the Offset method,
 for which the data points cannot move. Thus when the total uncertainty
 from both experimental and model sources is computed 
there is no great difference between these two aproaches.  

\begin{figure}[tbp]
\centerline{
\epsfig{figure=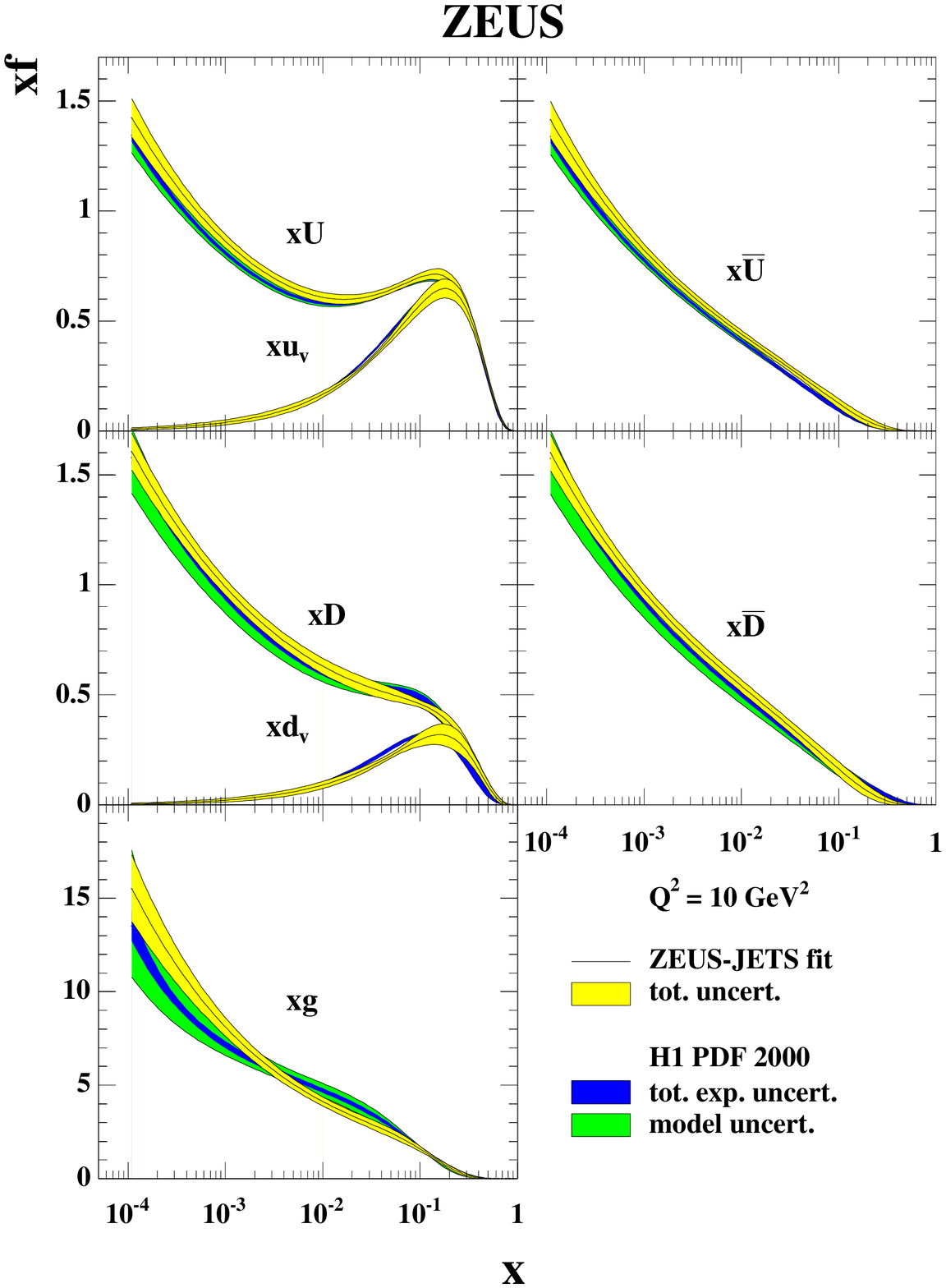,width=0.5\textwidth,height=7cm}
\epsfig{figure=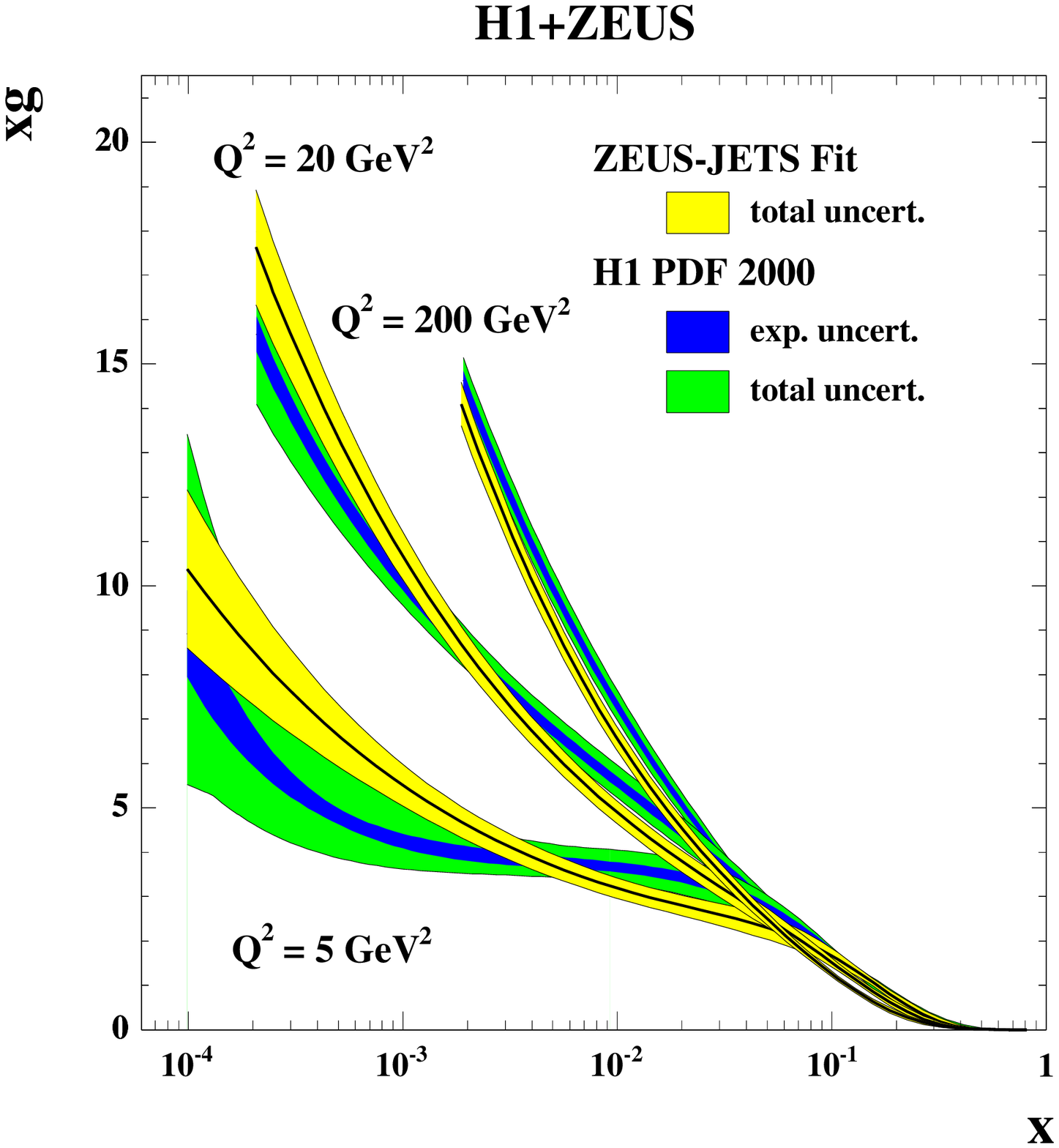,width=0.5\textwidth,height=6cm}}
\caption {Left plot: Comparison of PDFs from ZEUS and H1 analyses at $Q^2=10$GeV$^2$.
Right plot: Comparison of gluon from ZEUS and H1 analyses, at various $Q^2$. 
Note that the ZEUS analysis total uncertainty includes both experimental and 
model uncertainties.}
\label{fig:h1zeus}
\end{figure}

Fig.~\ref{fig:h1zeus} compares the results of the H1 and ZEUS analyses and
 illustrates the 
comparability of the ZEUS (Offset) total uncertainty  
estimate to the H1 (Hessian) experimental 
plus model uncertainty estimate. 
 Whereas the extracted PDFs are broadly compatible within errors, 
there is a 
noticeable difference in the shape of the gluon PDFs. This can be traced to small but significant differences in the $Q^2$ slope of low-$Q^2$ data.
Thus there could be an advantage in combining ZEUS and H1  
data into a single data set~\cite{combination}, not just in terms of reducing 
statistical errors, but also in reducing systematic errors by using each 
experiment to calibrate the other. 

\section{Combining ZEUS and H1 HERA-I data}
\label{sec:combine}
\begin{figure}[tbp]
\vspace{-2.0cm} 
\centerline{
\epsfig{figure=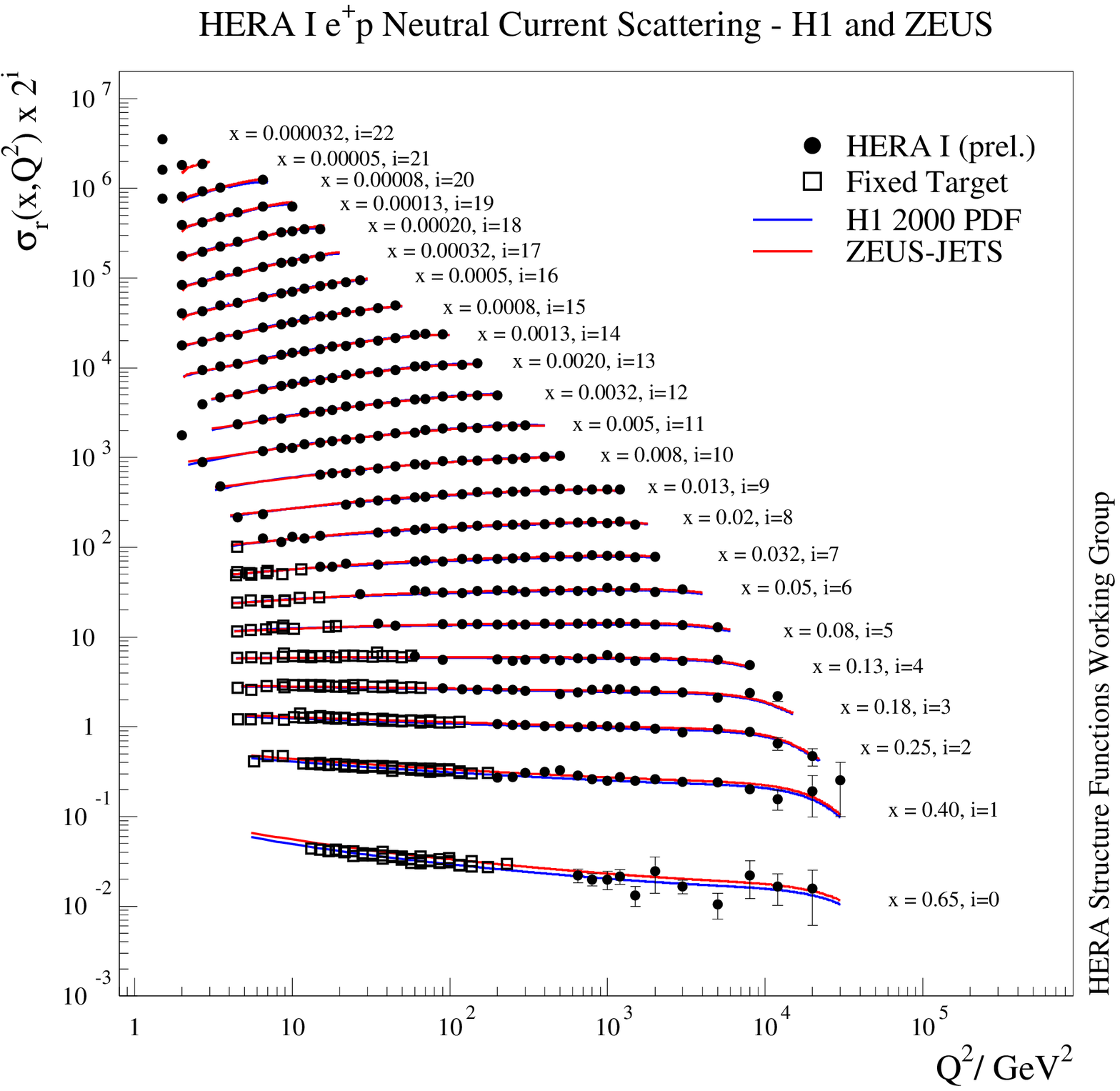,height=9cm}
\epsfig{figure=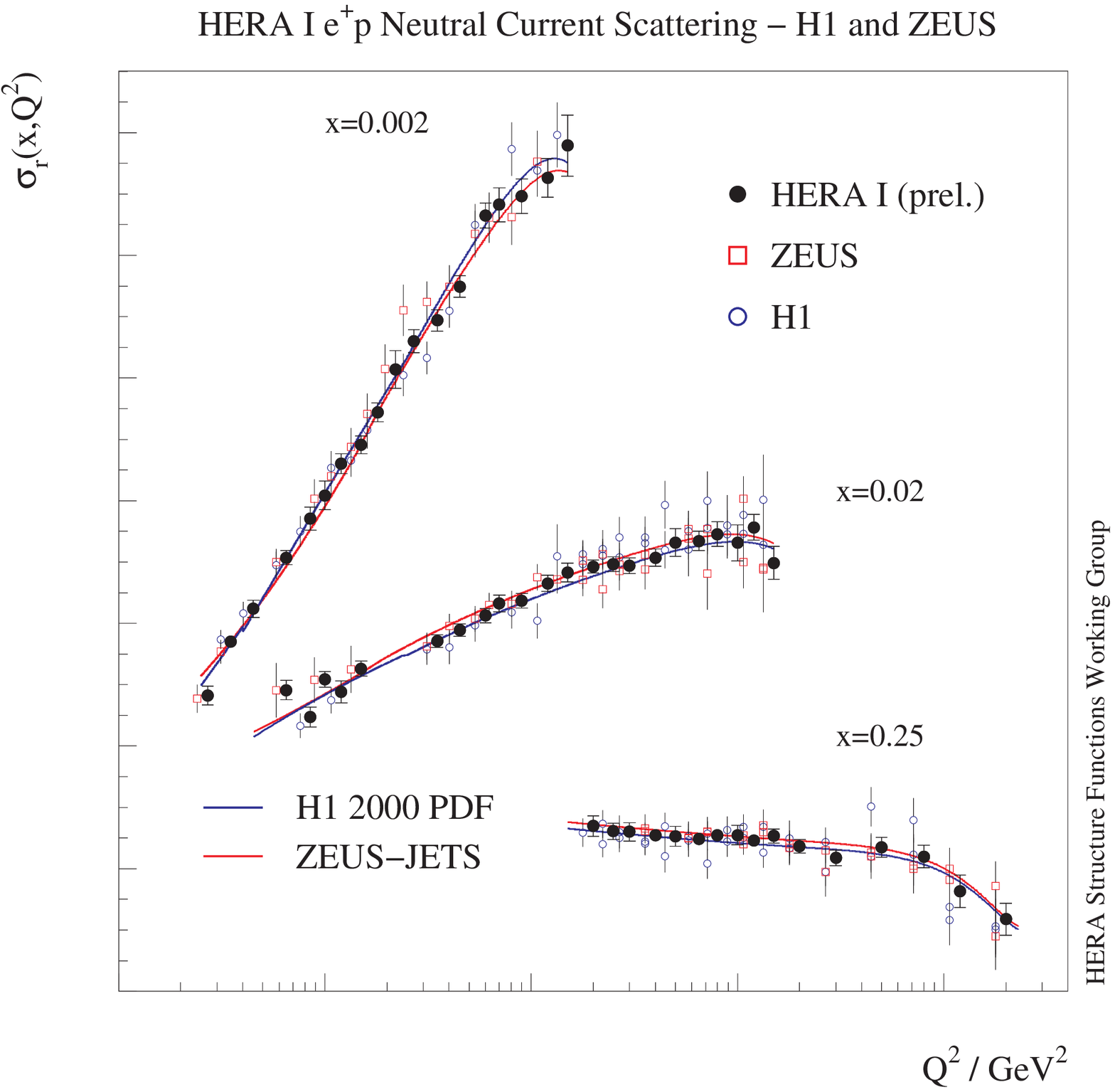,height=7cm}
}
\caption {Left hand side: HERA-I combined data on $\sigma_r$ as a function of $Q^2$ 
for NC $e^+p$ scattering, together with fixed target data, for $x$ bins across the whole 
measured kinematic plane. Right hand side: H1, ZEUS and HERA-I combined data on 
$\sigma_r$ for NC $e^+p$ scattering for low, middling and high-$x$. }
\label{fig:hera}
\end{figure}

Essentially, since ZEUS and H1 are measuring the same physics in the same 
kinematic region, one can try to combine them using a 'theory-free' 
Hessian fit in which the only assumption is that there is a true 
value of the cross-section, for each process, at each $x,Q^2$ point. 
The systematic uncertainty parameters, $s_\lambda$, of each experiment 
are fitted to determine the best fit to this assumption. 
Thus each experiment is calibrated to the other. This works well because the 
sources of systematic uncertainty in each experiment are rather different. 
Once the procedure has been performed the resulting systematic uncertainties 
on each of the combined data points are significantly smaller than the 
statistical errors. 
Fig.~\ref{fig:hera} shows the NC $e^+p$ reduced cross-sections from the HERA combination
and compares the individual H1 
and ZEUS  results with those of the combination so that the scale of the improvement 
can be appreciated..

\section{Adding exclusive jet cross-section data to PDF fits, and measurements of $\alpha_s(M_z)$}
\label{sec:jets}
The gluon PDF contributes only indirectly to the 
inclusive DIS cross sections, through the scaling violations. 
However it  makes a direct contribution to
jet cross sections through boson-gluon and quark-gluon 
scattering, so that measurements 
of these cross sections can constrain the gluon density.
Furthermore, the addition of the 
jet production data allows an accurate 
determination of $\alpha_s(M_Z)$ to be made in a simultaneous fit for $\alpha_s(M_Z)$ and 
the PDF parameters. 

In the ZEUS-JETS PDF fit, ZEUS neutral current $e^+p$ DIS inclusive 
jet cross sections  and direct 
photoproduction dijet cross sections have been used to 
constrain the gluon. 
The predictions for the jet cross sections were calculated to NLO in QCD 
using the programme of Frixione and Ridolfi~\cite{np:b507:315} 
for photoproduced dijets and 
{\sc Disent}~\cite{np:b510:503} for jet production in DIS. 
These calculations
are too slow to be used iteratively in the fit. 
Thus, they were used to compute LO and NLO weights, $\tilde{\sigma}$, 
which are independent of $\alpha_s$ and the 
PDFs, and are obtained by integrating the corresponding partonic hard cross 
sections\footnote{For the dijet photoproduction cross sections,
the weights also included the convolution with the photon PDFs.} in
bins of $\xi$ (the proton momentum fraction carried by
the incoming parton), $\mu_F$ (the factorisation scale) and
 $\mu_R$ (the renormalisation scale).
The predictions for the NLO QCD cross sections are then 
obtained by folding these weights with the PDFs and $\alpha_s$ according to the
formula
\begin{equation}
         \sigma = \sum_n \sum_a \sum_{i,j,k} 
                  f_a({\langle \xi \rangle}_i , {\langle \mu_F \rangle}_j)  
                  \cdot \alpha_s^n({\langle \mu_R \rangle}_k) 
                  \cdot \tilde{\sigma}^{(n)}_{a,\{i,j,k\}} ~,
\end{equation}
where the three sums run over the order $n$ in $\alpha_s$, the flavour $a$ of 
the incoming parton, and the indices ($i,j,k$) of the $\xi$, $\mu_F$ and 
$\mu_R$ bins, respectively. 
This procedure reproduces the NLO predictions to better than $0.5\%$.

The cross-section predictions for photoproduced jets are sensitive to the 
choice of the input photon PDFs. The AFG photon 
PDF~\cite{zfp:c64:621} was 
used in the fits, but in order to minimise sensitivity to this choice, the 
analysis was restricted to use only the `direct' photoproduction 
cross sections. These are defined by the cut $x^{\rm obs}_\gamma > 0.75$, 
where $x^{\rm obs}_\gamma$ is a measure of the fraction of the photon's 
momentum that enters into the hard scatter. 

\begin{figure}[tbp]
\vspace{-2.0cm} 
\centerline{
\epsfig{figure=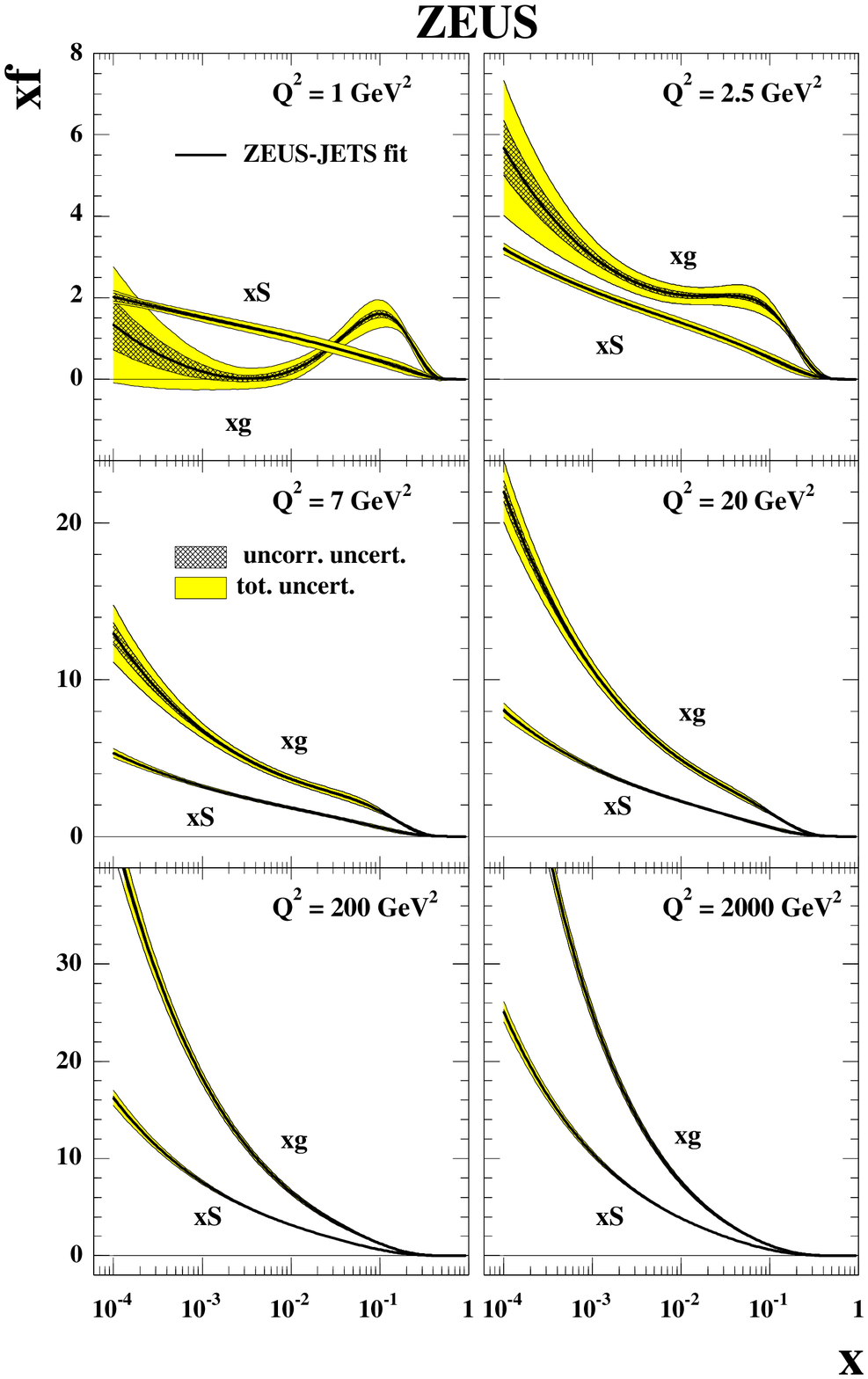,width=0.5\textwidth,height=7cm}
\epsfig{figure=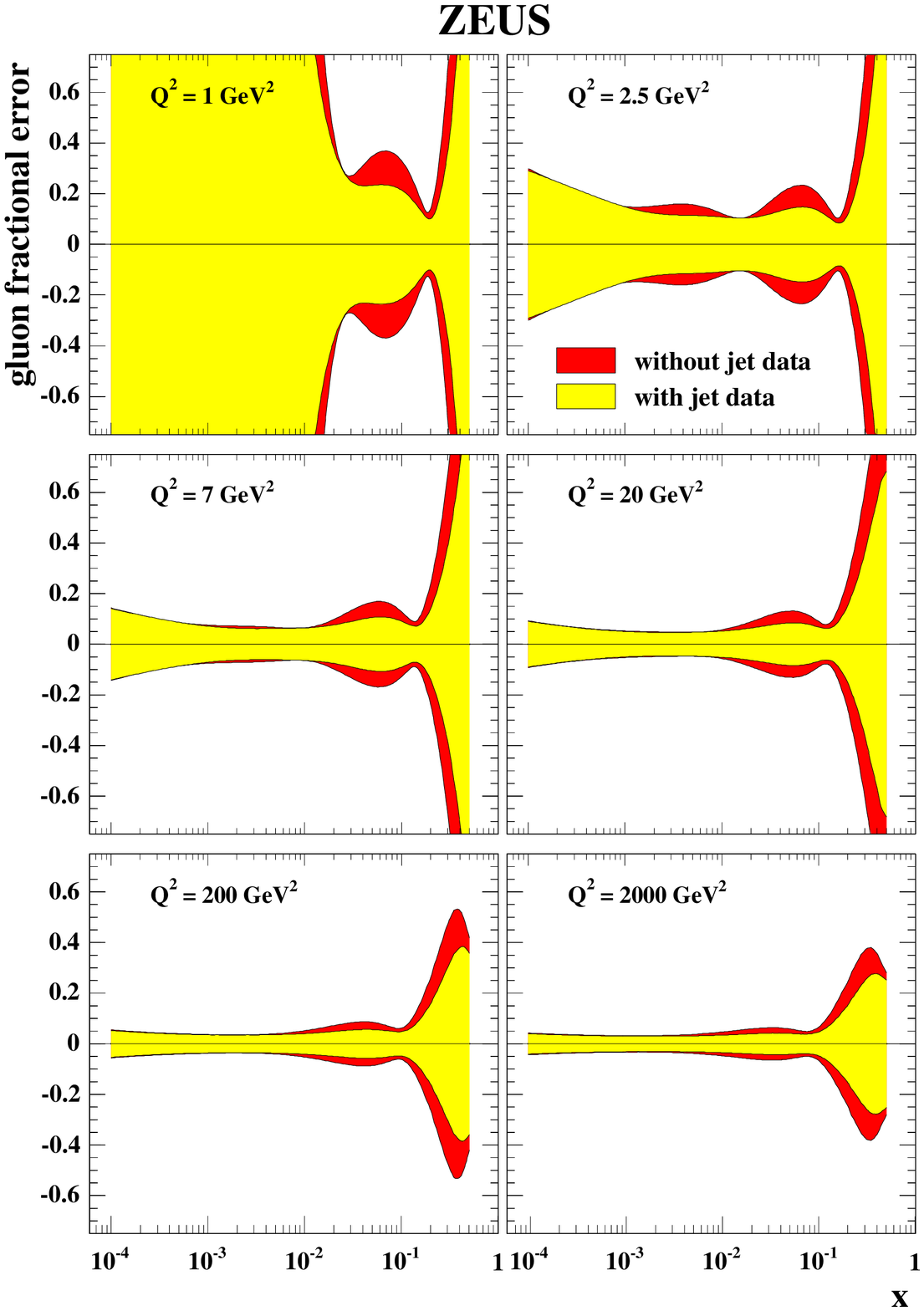,width=0.5\textwidth,height=6cm}
}
\caption {Left hand side: the sea and gluon PDFs from the ZEUS-JETS fit for various $Q^2$ 
values. 
Right hand side:The fractional error on the gluon PDF for various $Q^2$ values, for PDF 
fits made with and without including the jet data}
\label{fig:jets}
\end{figure}
Fig~\ref{fig:jets} shows that the jet data constrain the gluon mainly in the range  
$ 0.01 \leqsim \xi \leqsim 0.4$, although
the momentum sum-rule ensures that the indirect 
constraint of these data is still significant at higher $x$. 
The decrease in the uncertainty on 
the gluon distribution is striking; for example at $Q^2 = 7$~GeV$^2$ and 
$x = 0.06$ the uncertainty is reduced from $17\%$ to $10\%$. A similar decrease
in uncertainty by a factor of about two is found in this mid-$x$ range, over 
the full $Q^2$ range.

\begin{figure}[tbp]
\vspace{-0.5cm}  
\centerline{
\epsfig{figure=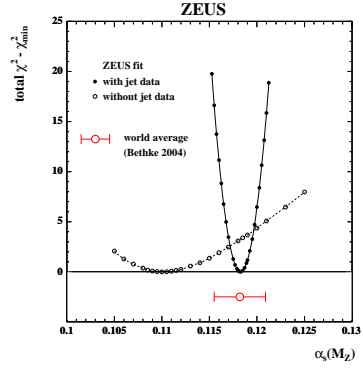,height=5cm}
}
\caption {The $\chi^2$ profiles vs. $\alpha_s(M_Z)$ for ZEUS PDF fits made with and without
 including the jet data }
\label{fig:chiprof}
\end{figure}
The value of $\alpha_s(M_Z)$
is fixed in most PDF fits but a simultaneous fit for $\alpha_s(M_Z)$ and the PDF parameters
can be made. Such fits to inclusive cross-section data do not yield accurate values of 
$\alpha_s(M_Z)$ because of the strong correlation between $\alpha_s(M_Z)$ and the gluon 
shape which comes from the DGLAP equations. However including jet data in the fit provides 
additional constraints. In the ZEUS-JETS fit with free $\alpha_s(M_Z)$ the value 
\[ \alpha_s(M_Z) = 0.1183 \pm 0.0027({\rm exp.}) 
\]
is obtained. Figure~\ref{fig:chiprof} illustrates  
the improved accuracy of the extraction of $\alpha_s(M_Z)$  
due to the inclusion of the jet data. The $\chi^2$ 
profile around the minimum is shown as a function of $\alpha_s(M_Z)$ for the 
ZEUS-JETS fit with $\alpha_s$ free, 
and a similar fit in which the jet data are not included.

\begin{figure}[tbp]
\vspace{-2.0cm} 
\centerline{
\epsfig{figure=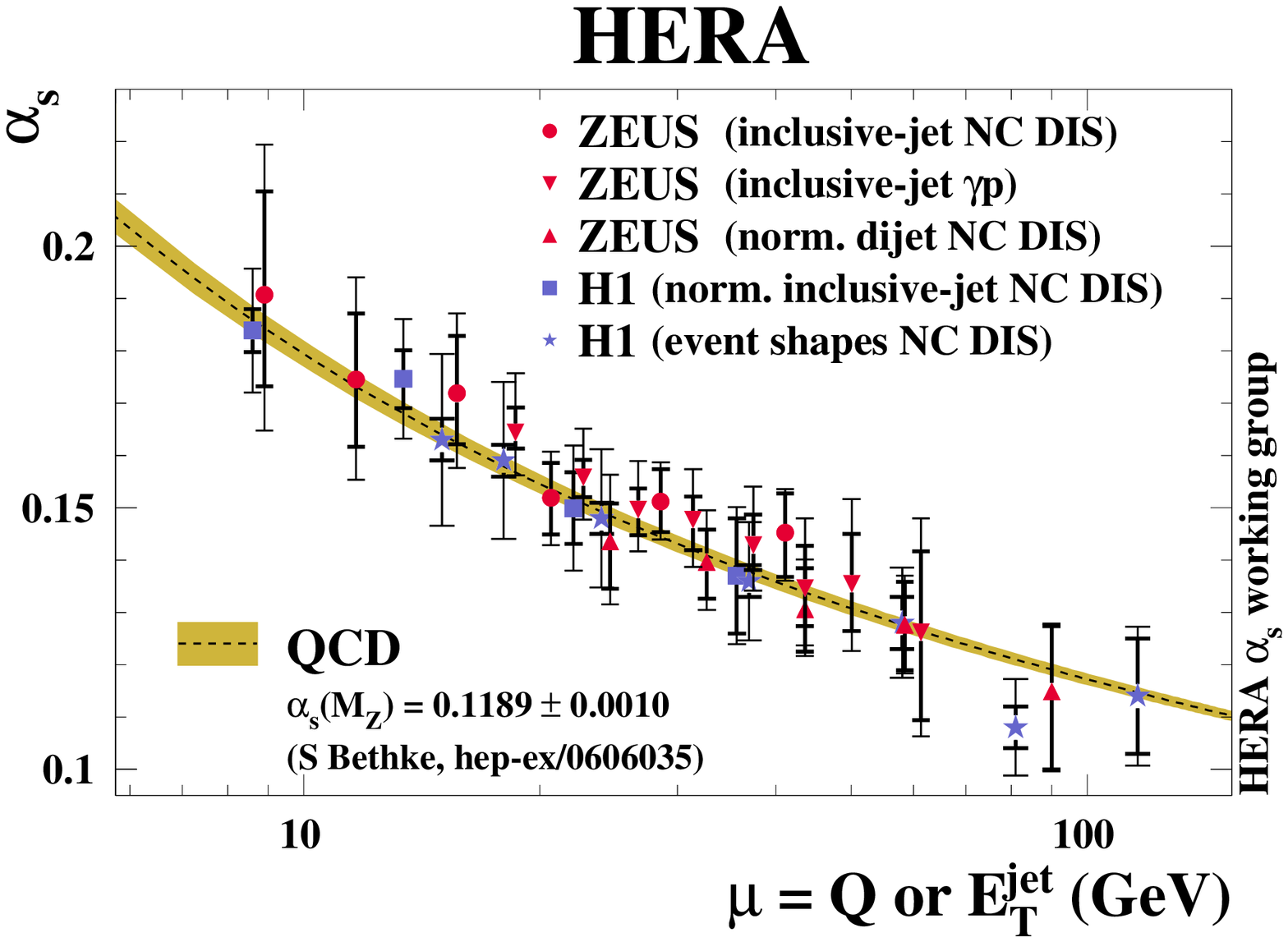,width=0.5\textwidth}
\epsfig{figure=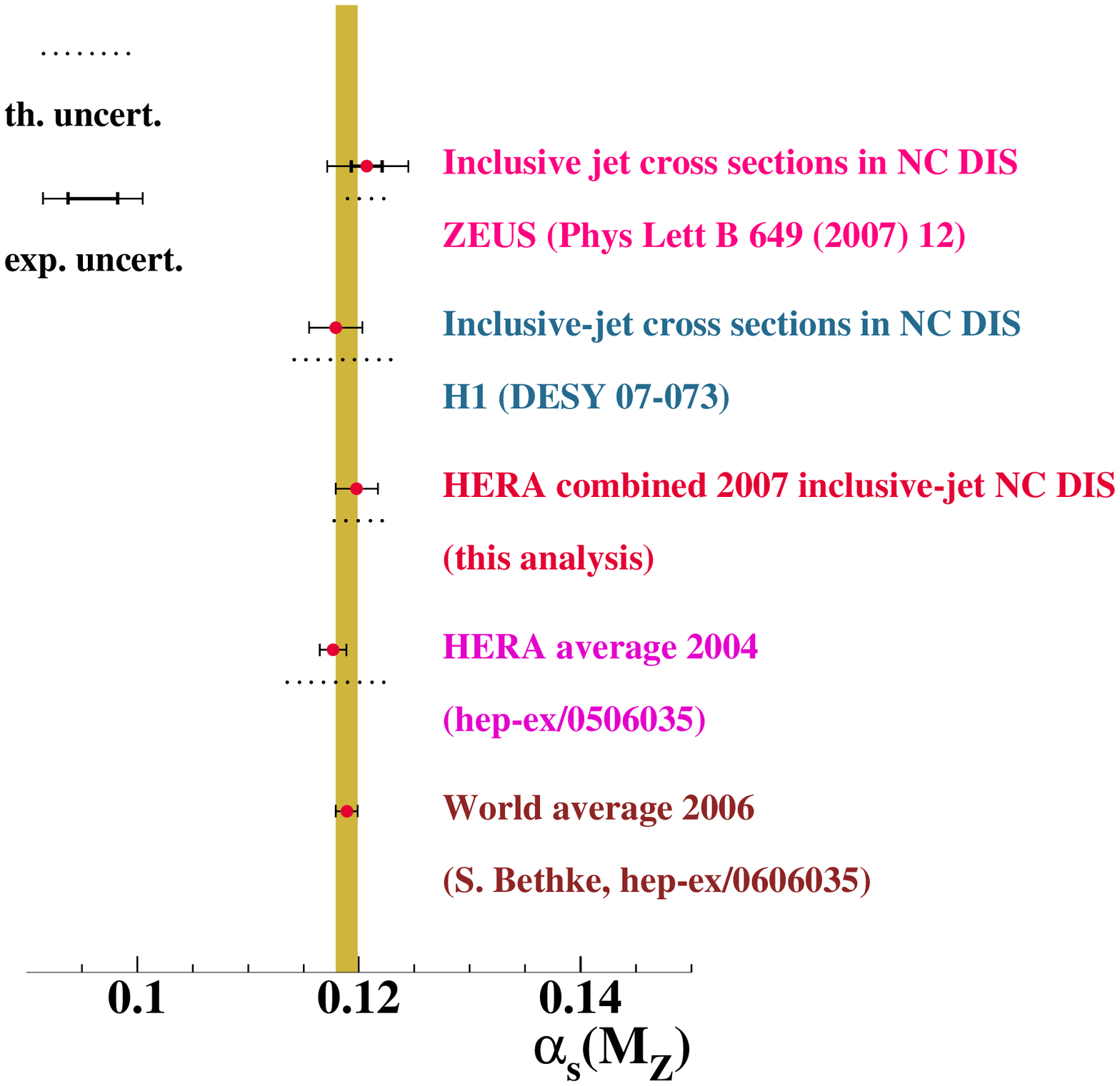,width=0.5\textwidth}}
\caption {Left hand side: $\alpha_s$ as a function of $Q^2$ determine from HERA data.
Right hand side: $\alpha_s(M_Z)$ values from H1, ZEUS and HERA combined determinations,
compared to the world average}
\label{fig:heraalf}
\end{figure}

There have also been accurate determinations of $\alpha_s(M_Z)$ using HERA jet 
data independent of PDF fits and a combined ZEUS and H1 $\alpha_s(M_Z)$ extraction has been
made~\cite{claudia}. Fig.~\ref{fig:heraalf} compares this combined value to those of the 
individual experiments and to the world average, and also illustrates the running of 
$\alpha_s$ with $Q^2$ as determined from the HERA experiments.  

\section{Adding HERA-II data to PDF fits}
\label{sec:heraii}
\begin{figure}[tbp]
\vspace{-1.0cm} 
\centerline{
\epsfig{figure=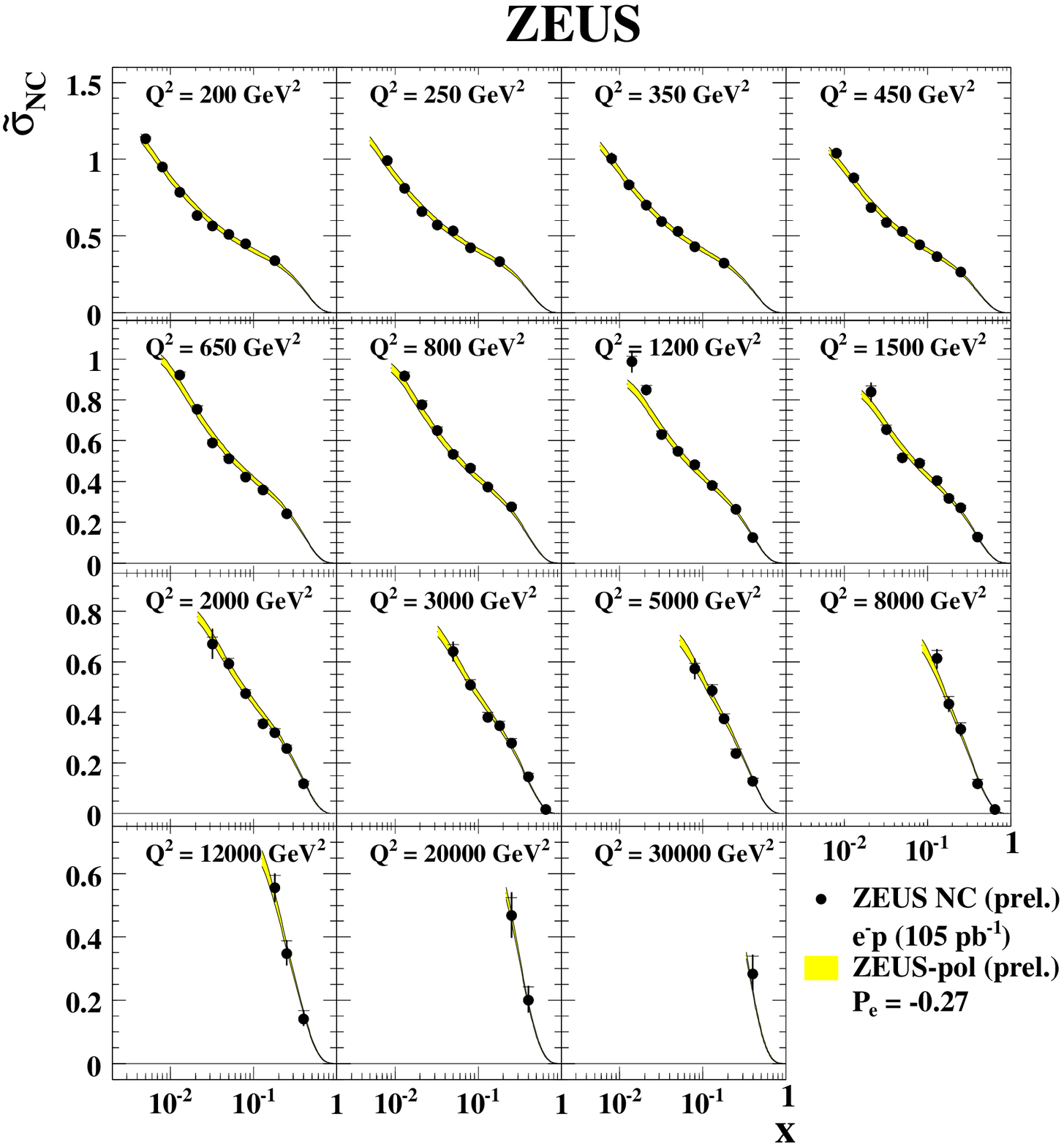,width=0.5\textwidth,height=6.5cm}
\epsfig{figure=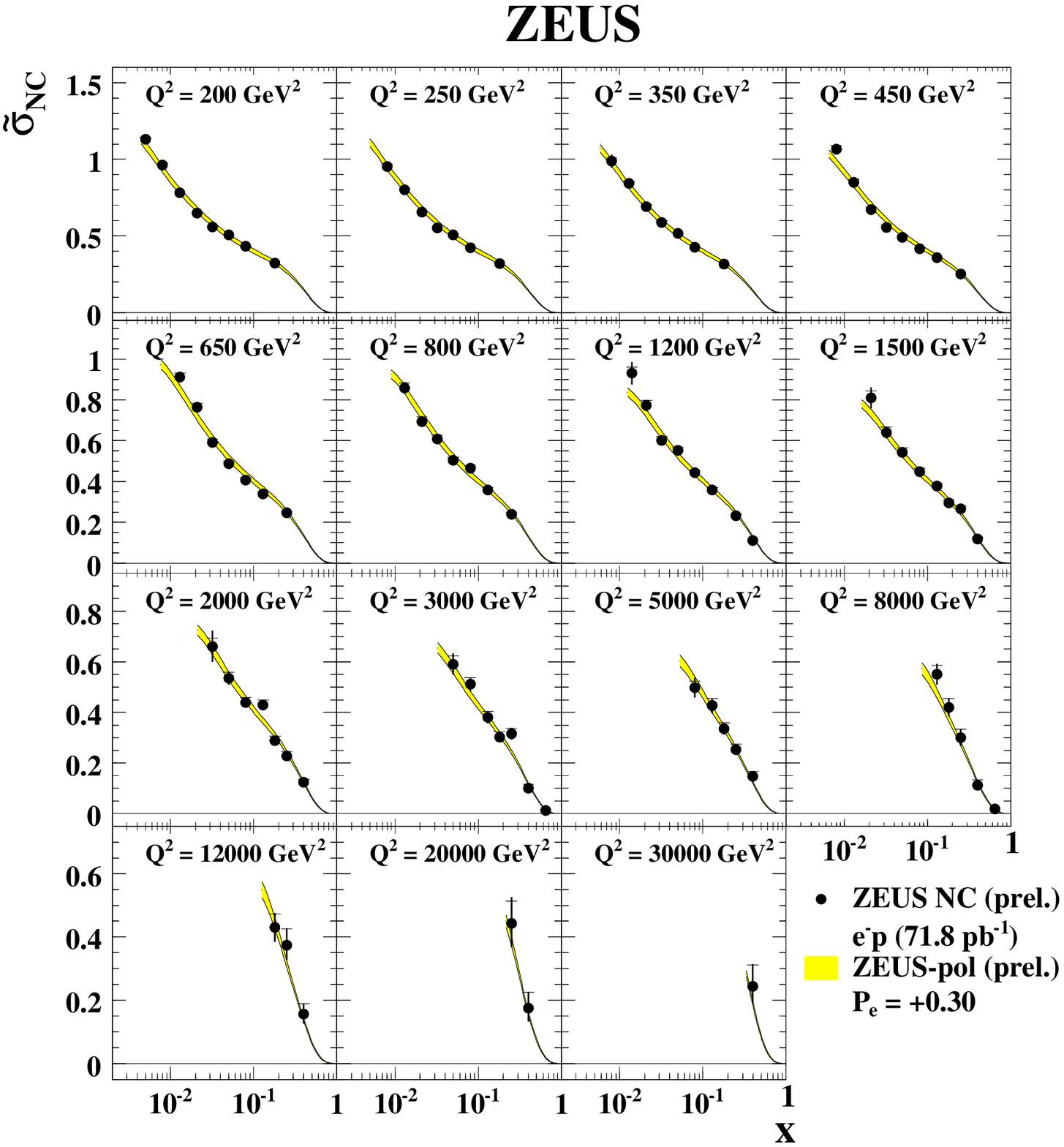,width=0.5\textwidth,height=6.5cm}}
\caption {ZEUS NC $e^-p$ data from HERA-II running with polarised beams. 
The predictions of the ZEUS-pol fit are superimposed}
\label{fig:zeuspol}
\end{figure}

\begin{figure}[bp]
\vspace{-0.65cm} 
\centerline{
\epsfig{figure=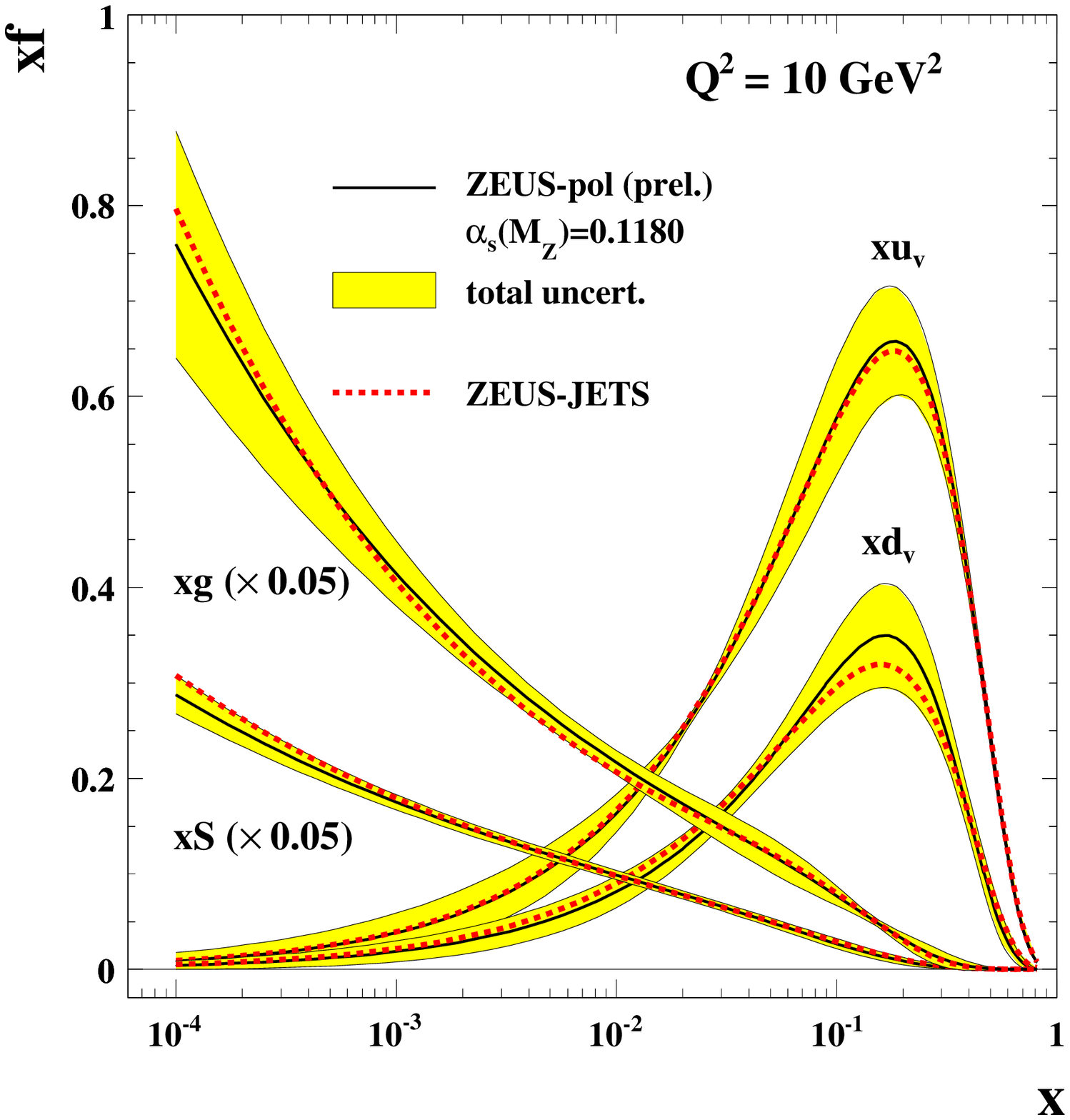,width=0.33\textwidth,height=5cm}
\epsfig{figure=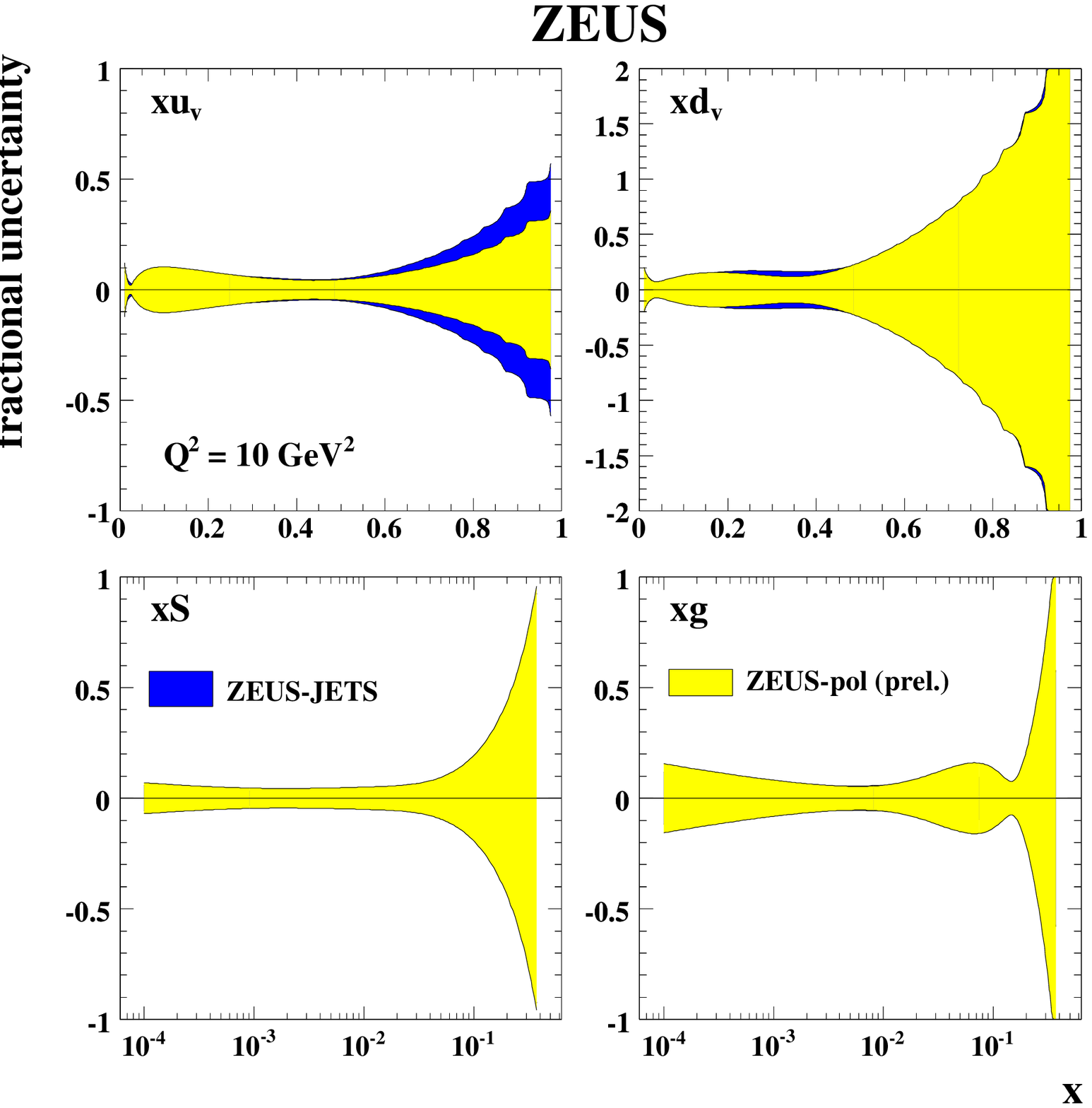,width=0.33\textwidth,height=5cm}
}
\caption {Left hand side: The PDFs extracted from the ZEUS-pol fit and 
their uncertainties, compared to 
the ZEUS-JETS fit at $Q^2=10$GeV$^2$. Right hand side: The fractioanl 
uncertainties of the ZEUS-pol PDFs 
compared to those of the ZEUS-JETS PDFs }
\label{fig:zpPDF}
\end{figure}

The determinations of the valence PDFs from HERA-I data are not as accurate as those from 
global fits, but this is rapidly improving with the addition of HERA-II data. 
Fig~\ref{fig:zeuspol} shows ZEUS $e^-p$ NC data from the 2004-6 running period 
with polarised beams. There are $105 pb^{-1}$ of negatively polarised, $P_e=-0.27$, data
 and $71.8 pb^{-1}$ of positively polarised, $P_e=+0.30$, data. 
This data, and $e^-p$ CC data from 2004-5, have been input to the 
ZEUS-JETS fit analysis framework and this new fit is called the ZEUS-pol 
fit~\cite{zeuspol}. The polarization of the data has been exploited to measure the 
neutral current vector and axial vector couplings~\cite{zeuspol}. 
The results of this ZEUS-pol fit are superimposed on the data in Fig~\ref{fig:zeuspol}.
The PDFs extracted from the ZEUS-pol fit are compared to those of the 
ZEUS-JETS fit in Fig~\ref{fig:zpPDF}. The central 
values of the fit  are very compatible with the 
ZEUS-JETS fit, and the $u$-valence quark uncertainty is reduced significantly 
at large $x$.
The improvement is mostly in the $u$-valence quark at present because 
the $e^-p$ data are
 $u$ quark dominated at large $x$. We can expect improvements in 
the $d$-valence 
distribution when the final $e^+p$ CC HERA-II data become available.  

\section{The future}
\label{sec:future}
\begin{figure}[tbp]
\vspace{-1.5cm} 
\centerline{
\epsfig{figure=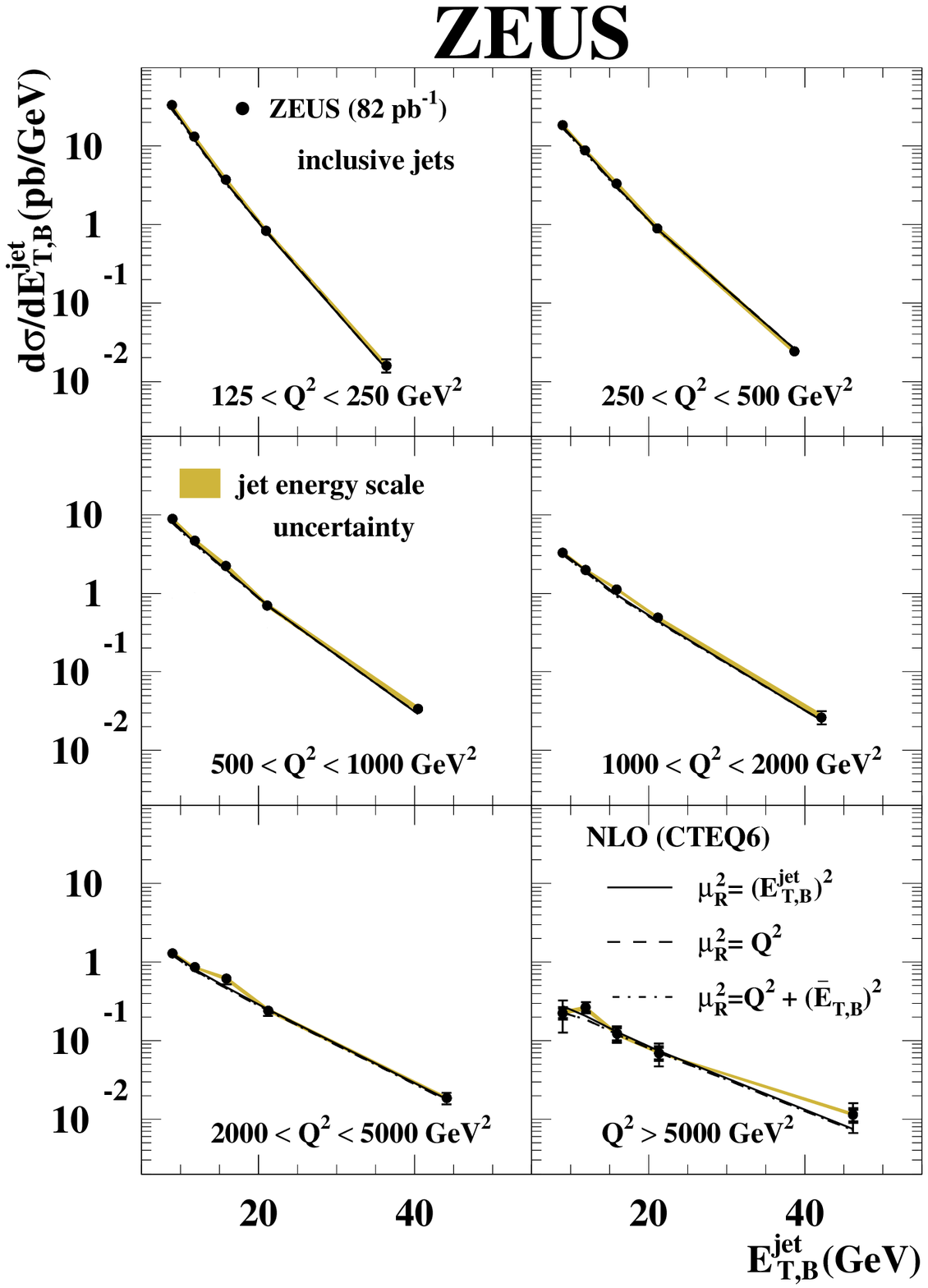,width=0.33\textwidth,height=6cm}
\epsfig{figure=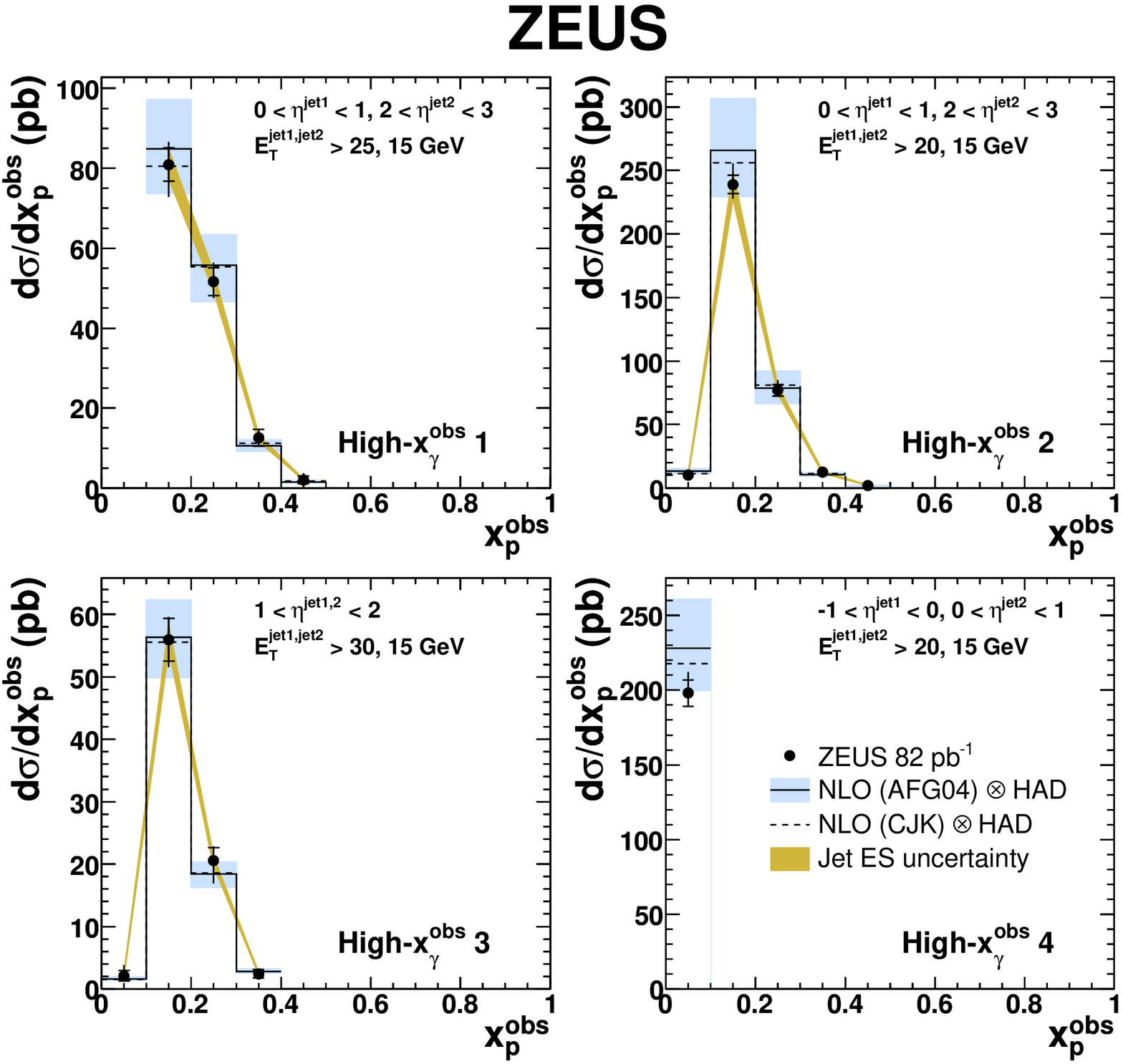,width=0.33\textwidth,height=6cm}
\epsfig{figure=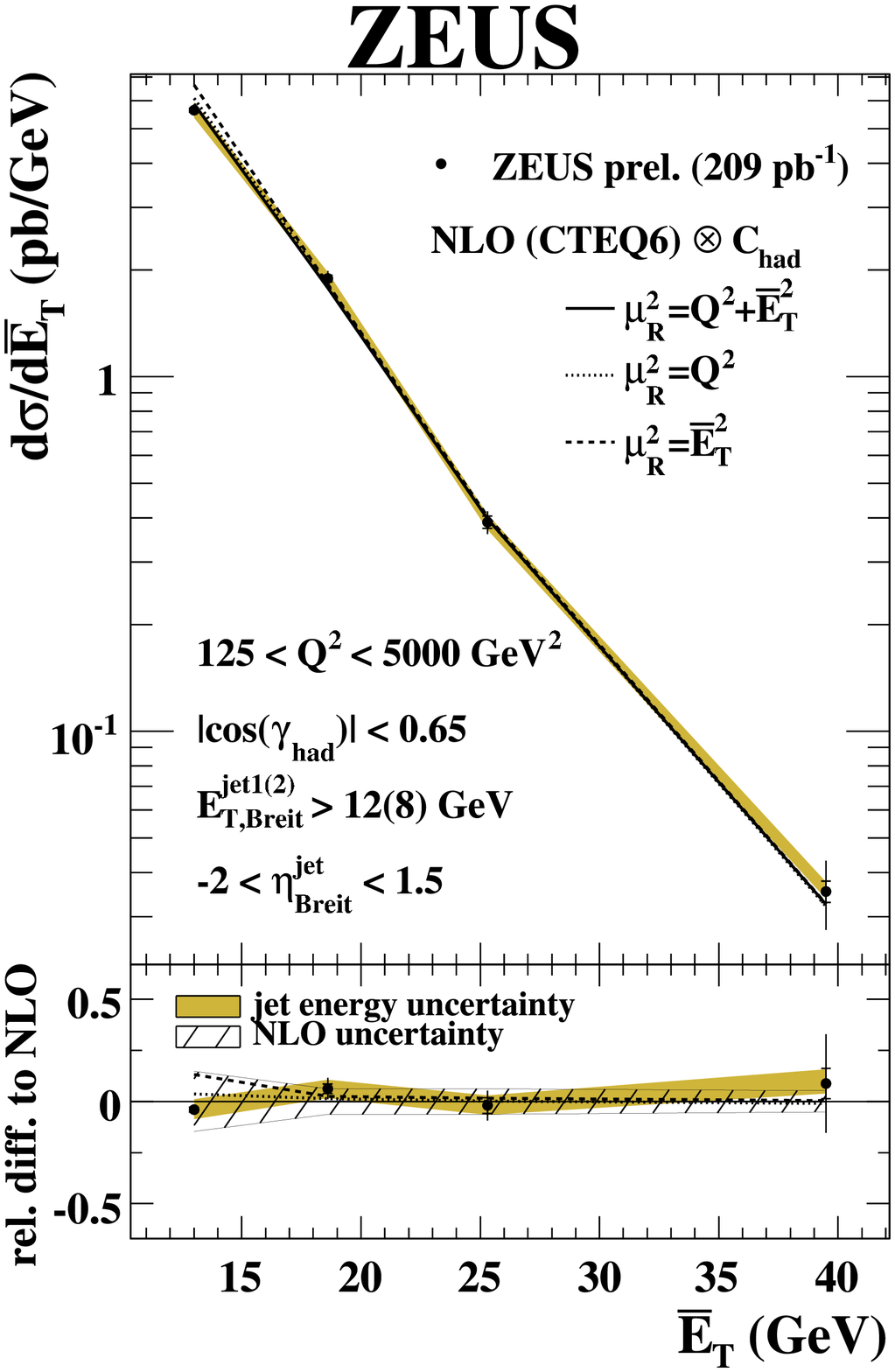,width=0.33\textwidth,height=6cm}}
\centerline{
\epsfig{figure=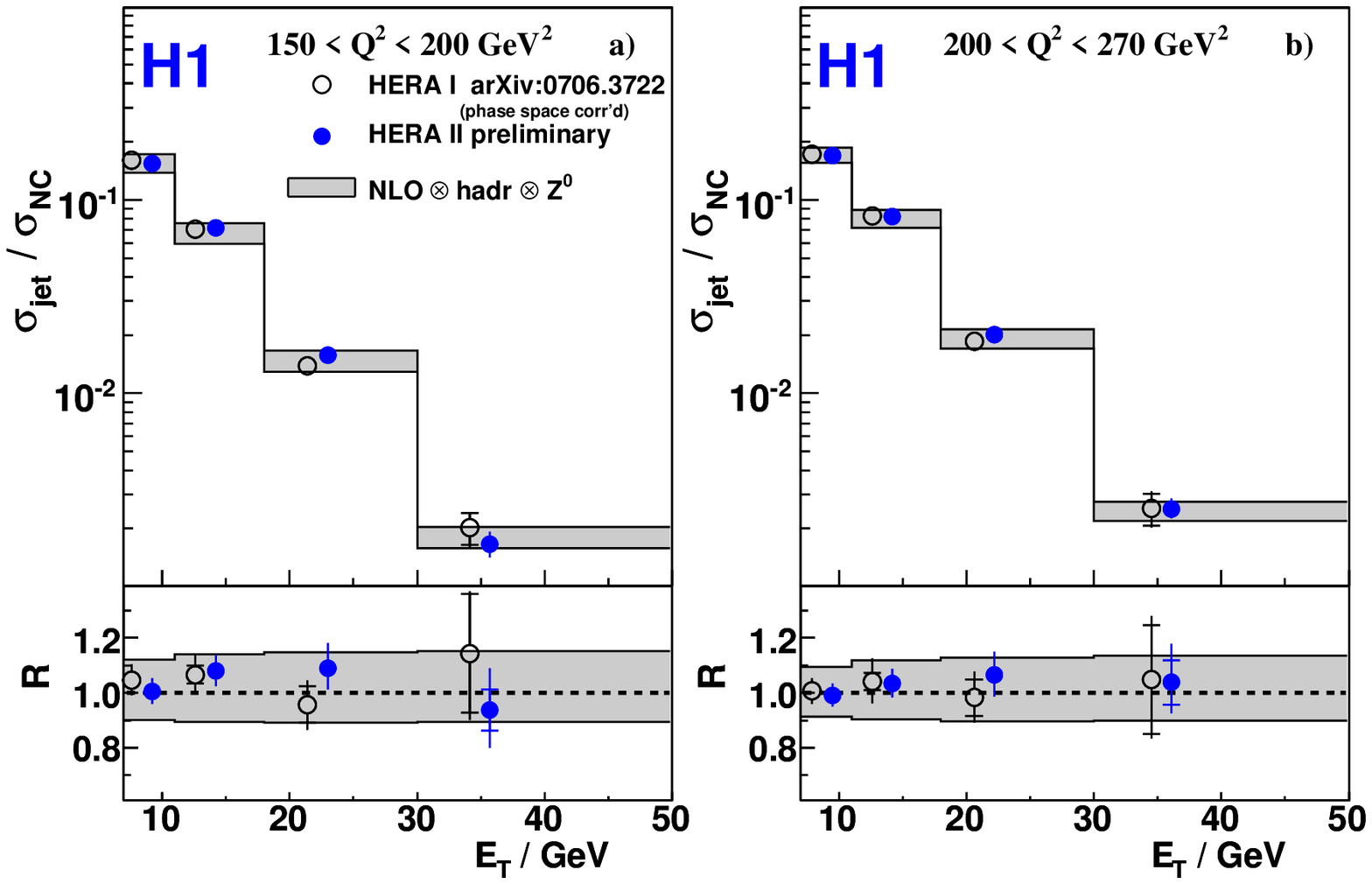,width=0.33\textwidth}
\epsfig{figure=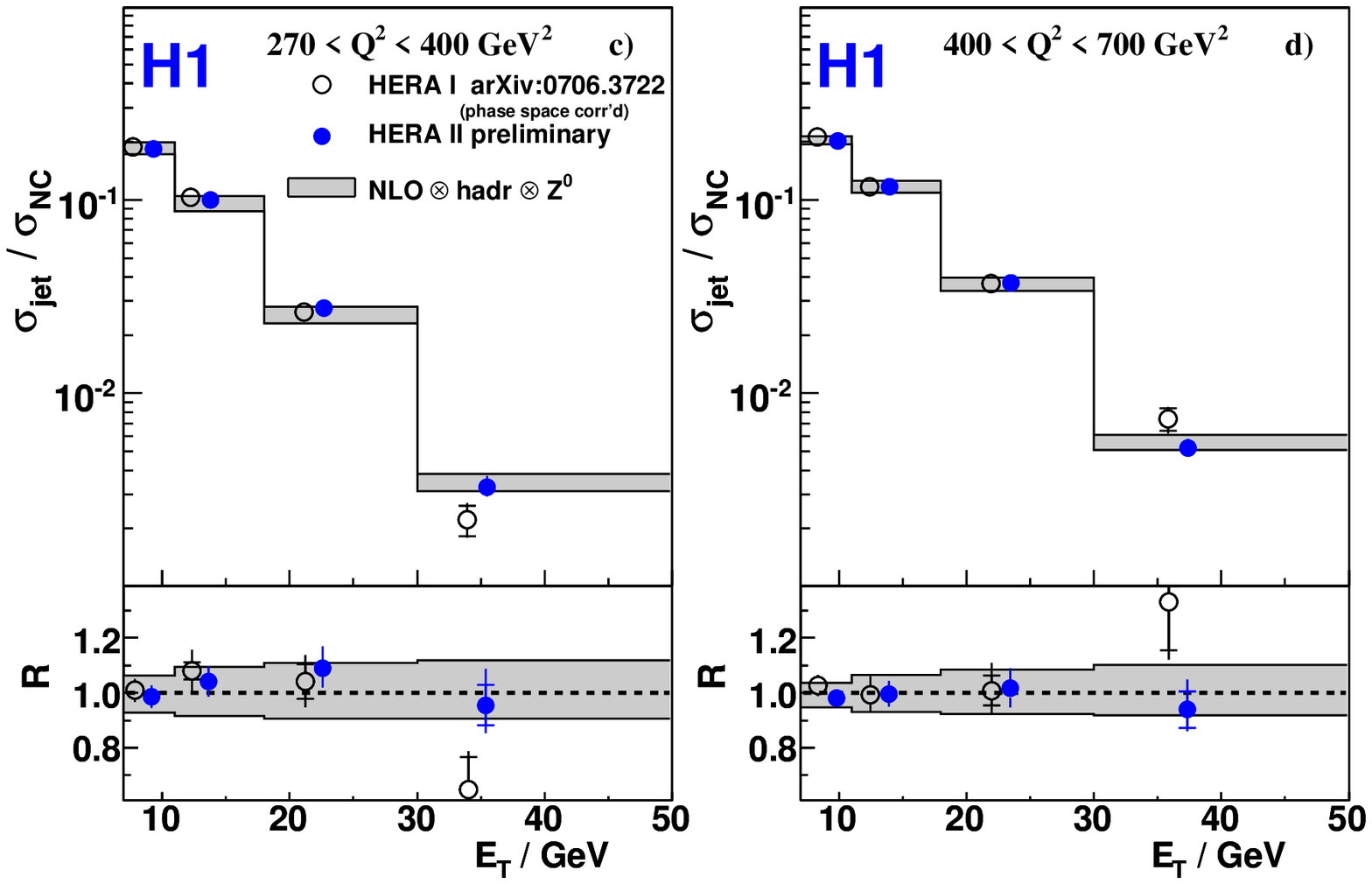,width=0.33\textwidth}
\epsfig{figure=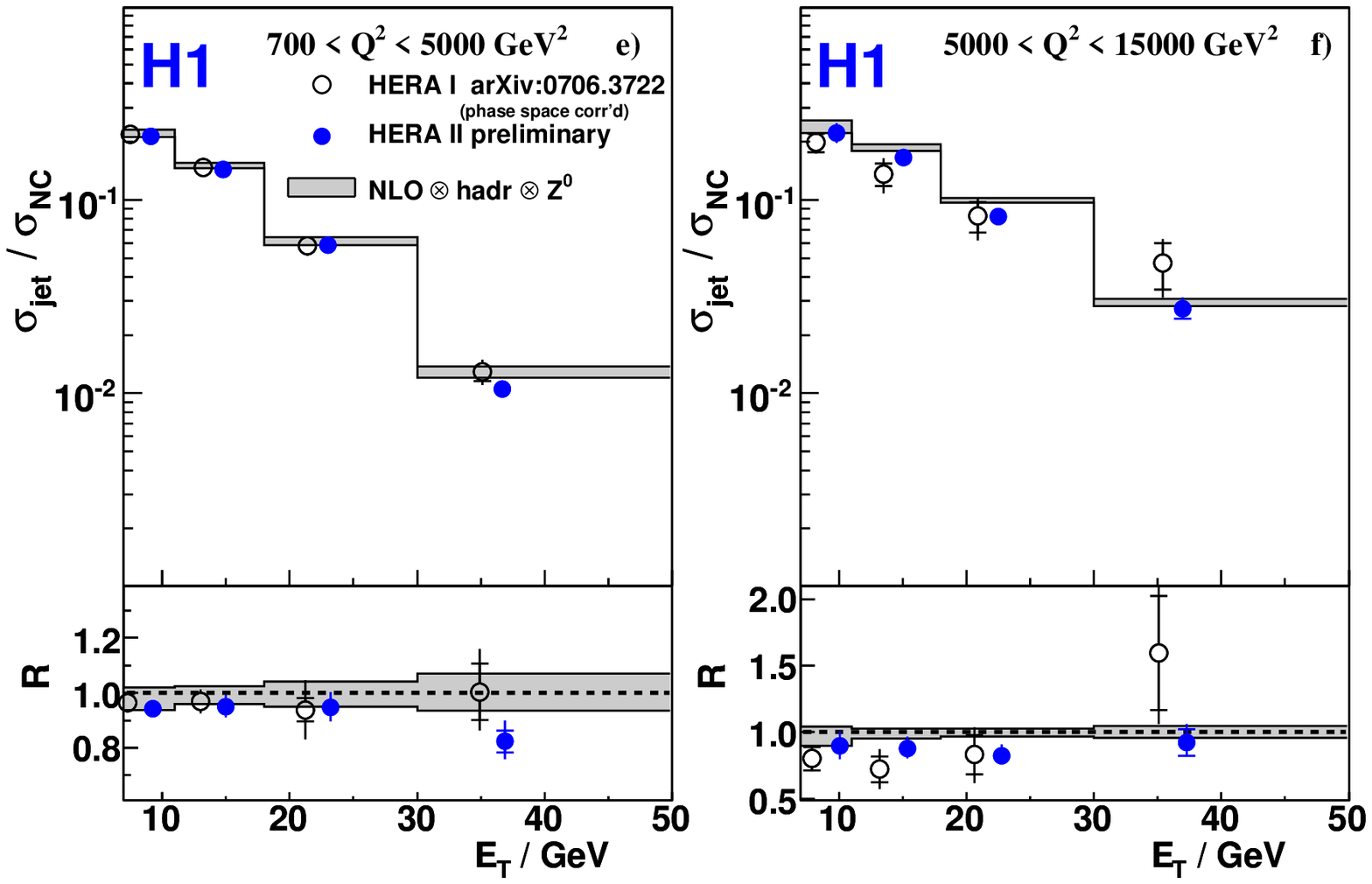,width=0.33\textwidth}}
\caption{Upper plots: ZEUS jet data from; (left) inclusive DIS (HERA-I, $82pb^{-1}$), 
(middle) di-jet photoproduction (HERA-I, $82pb^{-1}$), 
(right) di-jet DIS (HERA-II, $209pb^{-1}$). Lower plots: H1 normalised inclusive 
jet cross-sections (HERA-I and HERA-II, $320pb^{-1}$)
}  
\label{fig:morejets}
\end{figure}

We conclude with a look to the future. HERA data will continue to improve our knowledge 
of PDFs for the next few years.
Firstly, there is more jet data both from HERA-I~\cite{desy06128,desy07092,desy07073} 
and from $\sim 500 pb^{-1}$ of HERA-II~\cite{zeusprel07005,h1prel07131}
analyses, as shown 
in Fig~\ref{fig:morejets}. 
Inputting these data should improve determinations of the high-$x$ gluon. 

It is also interesting to investigate the low-$x$ gluon,
where the theoretical formalism of the NLO DGLAP equations may need
extending to account for $ln(1/x)$ resummation ~\cite{tw,abf,ccss} or even non-linear 
terms~\cite{cgc}. Fig.~\ref{fig:jets} shows the gluon and the sea PDFs predicted by the 
ZEUS-JETS fit. For $Q^2 \geqsim 7$GeV$^2$ the gluon PDF is larger than and steeper than 
the sea PDF, but for lower $Q^2$ it flattens and even becomes valence-like. This counter 
intuitive behaviour may come from the use of the DGLAP equations outside their region of 
applicability. At low $x$ the form of the DGLAP
equations is such that one has $F_2 \sim xq$ and $\frac{dF_2}{d \ln Q^2} \sim P_{qg} xg$. 
The determination of the gluon distribution is coming from the measurement of the scaling 
violations, $dF_2/d\ln Q^2$, but these may be determined by 
either the gluon density or the splitting function. Thus the odd behaviour of the gluon 
may in fact derive from use of an incorrect splitting function. The use of a calculation 
of the low-$x$ splitting functions which includes $ln(1/x)$ resummation results in a 
steeper gluon PDF~\cite{tw}. To settle these ambiguities definitively 
we need a measurement of the gluon density at small $x$ which does not derive from
the scaling violations of $F_2$, for example a  measurement of $F_L$ or $F_2^{c\bar c}$,
$F_2^{b\bar{b}}$ .

\begin{figure}[bp]
\vspace{-2.5cm} 
\centerline{
\epsfig{figure=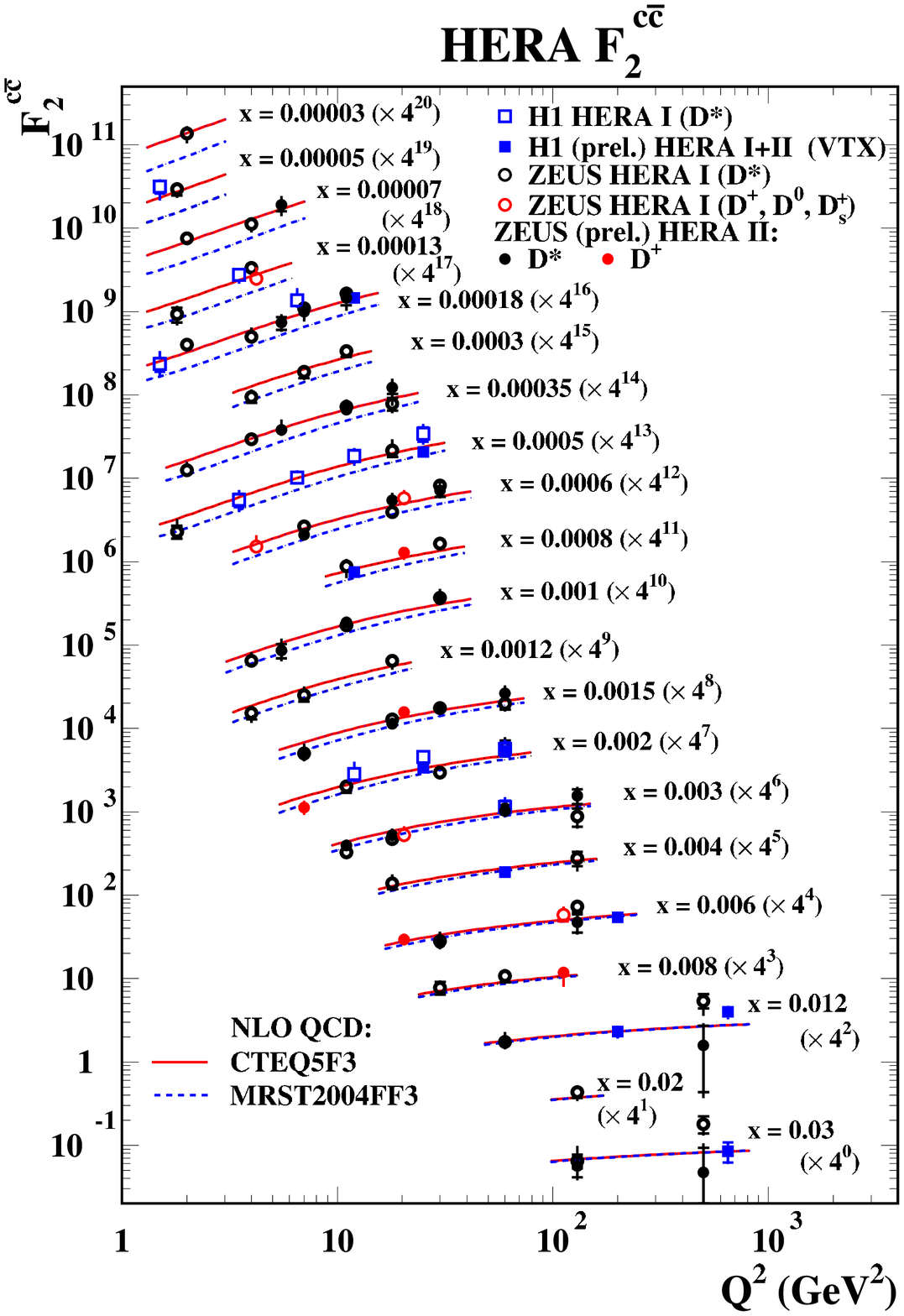,width=0.5\textwidth,height=9.5cm}
\epsfig{figure=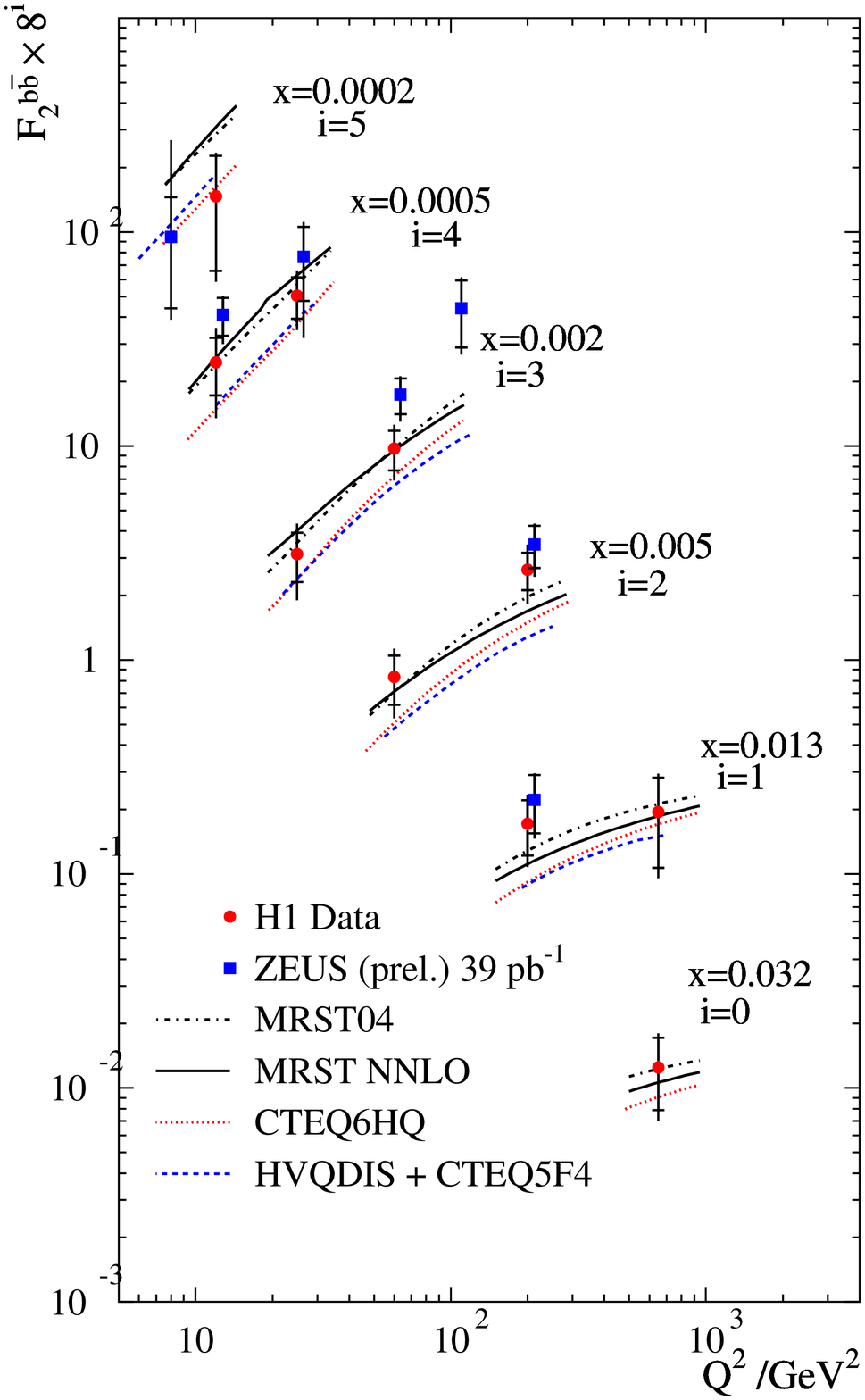,width=0.5\textwidth,height=8cm}}
\caption {Heavy quark production data compared to predictions from different PDF sets using 
different heavy quark production schemes. 
Left: $F_2^{c\bar{c}}$ data from HERA. Right:$F_2^{b\bar{b}}$ data from HERA. }
\label{fig:charm}
\end{figure} 
So far the addition of charm data to PDF fits has made little impact~\cite{amcscharm} 
but there is new 
data on $F_2^{c\bar{c}}$ from ZEUS, using $D$ production from $82pb^{-1}$ of HERA-I 
running~\cite{desy07052}, and using $D^*$ (and $D$) production from $162pb^{-1}$~\cite{eps106} 
(and $135pb^{-1}$~\cite{eps107}) of HERA-II running. 
There is also H1 data on $F_2^{c\bar{c}}$ from
$54pb^{-1}$~\cite{h1prel07171} of HERA-II data and these have been averaged together with 
the HERA-I data.  Both collaborations have also 
extracted $F_2^{b\bar{b}}$, H1 using the same data sample as for their charm extraction and ZEUS
using $39pb^{-1}$ of HERA-II data~\cite{eps108}. These data are shown in Fig.~\ref{fig:charm}. 
In principle heavy quark data should give information on the gluon distribution since 
heavy quarks are generated by $g\to c\bar{c}$ and $g\to b\bar{b}$. 
However, at the present time there is some theoretical disagreement about heavy quark 
production schemes~\cite{thompson,tung,hqnew} such that these data may tell us more 
about the correct treatment of heavy quarks than about PDFs.

\begin{figure}[tbp]
\vspace{-2.0cm} 
\centerline{
\epsfig{figure=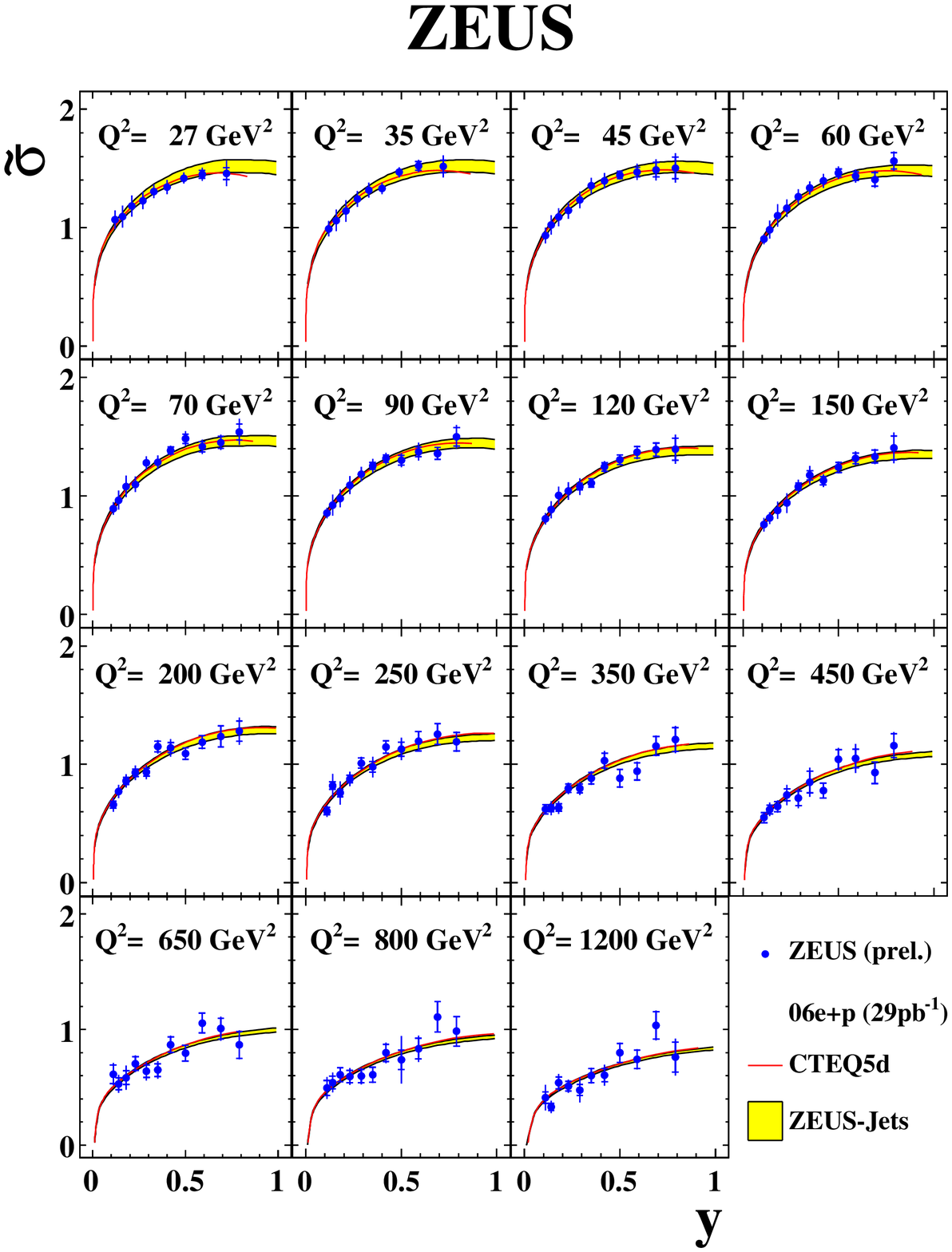,width=0.33\textwidth,height=7cm}
\epsfig{figure=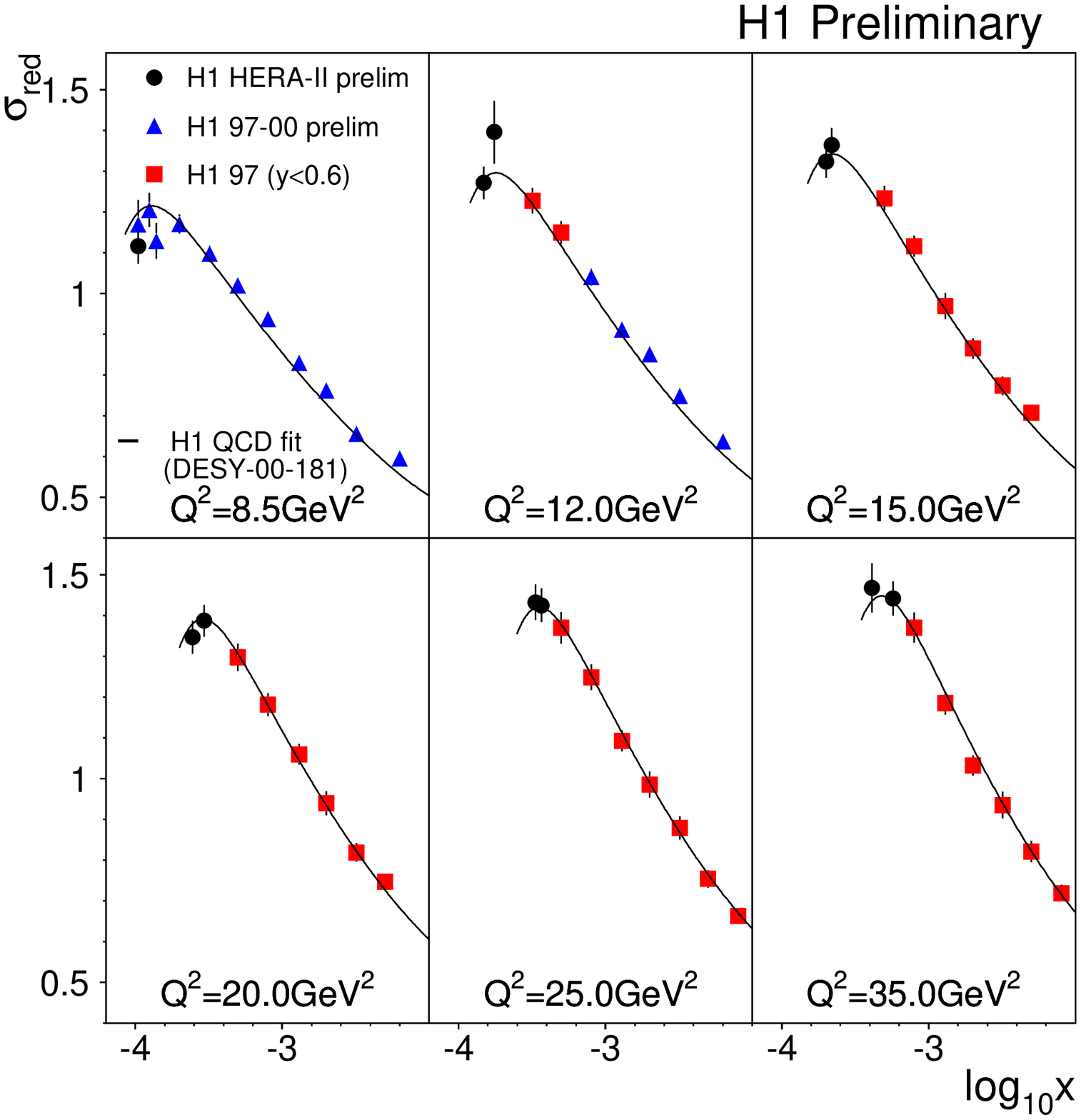,width=0.33\textwidth,height=6cm}
\epsfig{figure=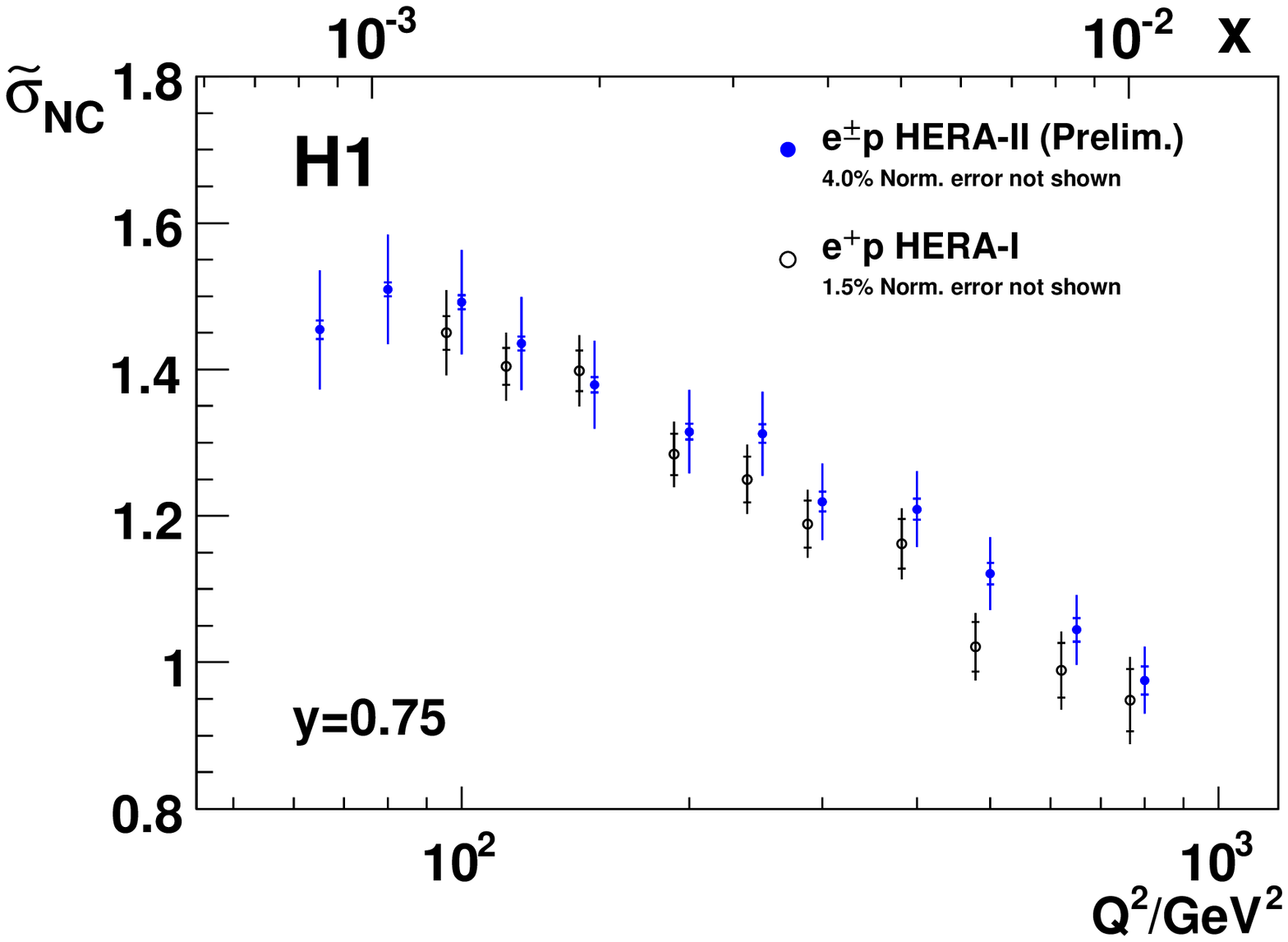,width=0.33\textwidth,height=5cm}
}
\caption {HERA data at high $y$ from: (left) ZEUS HERA-II, (middle) H1 HERA-I at low $Q^2$ and 
(right) H1 HERA-II at high-$Q^2$.}
\label{fig:highy}
\end{figure}

The structure function $F_L$ depends strongly on the gluon~\cite{amcsfl}. A model 
independent measurement of 
$F_L$ requires data at different beam energies so in 2007 HERA was 
run at proton beam energies $460$ GeV and $575$ GeV. $F_L$ only makes strong contributions to the cross-section 
at high 
$y$, and measurements at high-$y$ require technically challenging identification of low energy 
scattered electrons. Both collaborations have been preparing for this challenge by extending 
their measurement capabilities to high $y$ using the nominal energy HERA-I and  HERA-II running.
Fig~\ref{fig:highy} shows data at high-$y$ from ZEUS HERA-II 2006 running~\cite{eps78}, 
and from H1 HERA-I running, at low-$Q^2$~\cite{h1prel07042}, and HERA-II running,
at high-$Q^2$~\cite{h1prel07144}. These data not only pave the way for 
measurement of $F_L$, they are also interesting in their own right since they access 
a new kinematic regime. Thus we look forward to exciting new information on hadron structure
 from these measurements in the near future.

\bibliographystyle{heralhc} 
{\raggedright
\bibliography{heralhc}
}
\end{document}